\documentclass[12pt]{article}
\usepackage{hyperref}
\usepackage{amsmath, amsfonts, amssymb, mathrsfs}
\usepackage{dcolumn}
\usepackage{caption}
\usepackage{subcaption}
\usepackage{filemod}
\usepackage{natbib}
\usepackage[ruled, vlined]{algorithm2e}
\usepackage{floatrow}
\usepackage{setspace}

\usepackage{graphicx}
\graphicspath{{./figures/}}

\title{Active Learning for Deep Gaussian Process Surrogates}
\author{Annie Sauer\thanks{Corresponding author: Department of Statistics, 
	Virginia Tech, {\tt anniees@vt.edu}} \and Robert B. Gramacy\thanks{Department 
	of Statistics, Virginia Tech} 
	\and David Higdon\footnotemark[2]}
\date{\today}

\begin{document}

\maketitle
\bigskip

\begin{abstract} 
Deep Gaussian processes (DGPs) are increasingly popular as predictive models
in machine learning (ML) for their non-stationary flexibility and ability to
cope with abrupt regime changes in training data.  Here we explore DGPs as
surrogates for computer simulation experiments whose response surfaces exhibit
similar characteristics.  In particular, we transport a DGP's automatic
warping of the input space and full uncertainty quantification (UQ), via a
novel elliptical slice sampling (ESS) Bayesian posterior inferential scheme,
through to active learning (AL) strategies that distribute runs non-uniformly
in the input space -- something an ordinary (stationary) GP could not do.
Building up the design sequentially in this way allows smaller training sets,
limiting both expensive evaluation of the simulator code and mitigating cubic
costs of DGP inference. When training data sizes are kept small through 
careful acquisition, and with parsimonious layout of latent layers, the 
framework can be both effective and computationally tractable.  Our 
methods are illustrated on simulation data and two real computer 
experiments of varying input dimensionality.  We provide an open source 
implementation in the \texttt{deepgp} package on CRAN.
\end{abstract}



\section{Introduction}

Computer experiments are common when field data/physical measurement is scarce
or non-existent.  Examples include interactions in economic systems
\citep{deissenberg2008eurace}, movement of satellites in low-Earth orbit
\citep{mehta2014modeling}, and dispersion of diseases through populations
\citep{fadikar2018calibrating}.  Simulation can provide insight into complex
processes that might otherwise be immeasurable, but it may come with high
computational costs (some runs taking days to produce output for a single
input).  In these cases, a surrogate model may be fit from a limited,
carefully designed simulation campaign.  If predictive evaluations from the
surrogate are fast enough, and come with appropriate acknowledgment of
uncertainty, they may be valuable as replacements for actual simulations at
untried input settings. The canonical surrogate is based on Gaussian processes
(GPs).

GPs are favored for their partially analytic inference and nonlinear
predictive capability, including closed form variance calculations, which can
be used to drive active learning (AL): the sequential design and build-up of a
surrogate through a virtuous cycle of data acquisition and fitting.  See texts
from \citet{santner2018design}, \citet{rasmussen2005gaussian}, and
\citet{gramacy2020surrogates} for a thorough review.  Despite their
popularity, the typical assumption of stationarity -- made primarily for
computational convenience -- compromises the GP's ability to accommodate
features common to many computer simulations, such as regime changes in
input--output dynamics.  Early solutions to this problem from the spatial
statistics \citep[e.g.,][]{sampson1992nonparametric,higdon1999non,
paciorek2003nonstationary,schmidt2003bayesian} and machine learning (ML) \
perspective \citep[e.g.,][]{rasmussen2000the,rasmussen2002infinite} focused on
low input dimension (e.g., longitude and latitude) and small training data
sizes.

More recent advances in non-stationary spatial modeling
\citep[e.g.,][]{bornn2012modeling,katzfuss2013bayesian} emphasize scaling-up
to larger training data sets, still privileging low-dimensional input spaces.
The computer surrogate modeling literature now has bespoke solutions which
work well in specific, modest-dimensional cases: tree--GP hybrids for when
non-stationarity manifests along coordinate directions
\citep{gramacy2008bayesian}; composite processes when it arises as changes in
amplitudes \citep{ba2012composite}; local data subsets when training data are
massive \citep[e.g.,][]{gramacy2015local,cole2021locally}.  Yet general purpose
design and modeling strategies for an expensive, limited simulation campaign
remain elusive.
  
We see promise in the form of deep GPs, first proposed in the ML
literature by \citet{damianou2013deep}.  The setup neatly synthesizes elements
of warping, latent inputs, and composition as earlier introduced in isolation
by authors of several of the papers cited above.  By warping
the input space through hidden Gaussian layers, effectively moving some
training samples closer and others farther apart, they achieve
non-stationarity even under otherwise conventional kernel structures.  Prowess
has been demonstrated on many classification tasks
\citep{damianou2013deep,fei2018active,yang2020bayesian}.  Application as
surrogates for simulation experiments is rather less well developed
\citep{radaideh2020surrogate}.

The form of the DGP likelihood makes direct inference impossible. Initial schemes
leveraged approximate variational inference \citep{damianou2013deep}, embracing
computational thrift at the cost of full UQ.  Recent advancements from 
ML, including expectation propagation \citep{bui2016deep}, doubly 
stochastic variational inference \citep{salimbeni2017doubly}, and stochastic
gradient Hamiltonian Monte Carlo \citep{havasi2018inference}, perform well
when data is abundant. This is a mismatch to typical surrogate modeling 
scenarios in which efforts to limit expensive campaigns require careful,
uncertainty-driven sequential design.  We thus depart from the canonical
inferential apparatus to favor Markov chain Monte Carlo (MCMC) posterior
sampling via a novel hybrid Gibbs--Metropolis and elliptical slice sampling
\citep[ESS;][]{murray2010elliptical} scheme.  This is more expensive, but adds 
value in straightforward implementation and remains tractable in typical 
surrogate modeling scenarios.  Our main goal in this paper is to demonstrate, 
and provide details for, effective application of DGP surrogates and 
AL for simulation experiments.

Some inroads have been made along these lines.  \citet{dutordoir2017deep} fit
DGPs to sequentially collected data, but acquisition criteria were not based
on the DGP fits. \citet{rajaram2021empirical} suggested a maximum variance
criterion, a strategy sometimes called ``active learning MacKay''
\citep[ALM;][]{mackay1992information}, with DGPs.  ALM is under-powered
compared to aggregate variance-based criteria that are more common in the
computer experiments literature.  We propose using integrated mean-squared
error (IMSE) and its ML cousin ``active learning Cohn''
\citep[ALC;][]{cohn1994neural}, which has better properties and empirical
performance under GPs \citep{sambu2000gaussian,gramacy2009adaptive}.  
\citet{hebbal2021bayesian} applied DGPs to Bayesian optimization (BO) via
expected improvement \citep{jones1998efficient}, an extension we defer to
future work.

The crux of the value of our contribution lies in combining the right DGP
model specification with an appropriate inferential scheme from the
perspective of UQ and AL for sequential design.  We advocate for a
simple, limited DGP apparatus in terms of number of layers/dimensionality, and
demonstrate that such setups outperform both ordinary GPs and deeper/more
complex DGPs on benchmark and real-data surrogate modeling and design
exercises.  We furnish an \textsf{R} package called \texttt{deepgp} on 
CRAN, implementing every feature discussed in this
manuscript.   Code for all examples resides in a public git
repository (\url{https://bitbucket.org/gramacylab/deepgp-ex/}).

The remainder of the paper is outlined as follows.  Section \ref{sec:review}
reviews GPs, sequential design/AL criteria, and DGPs. Section
\ref{sec:modeling} details our proposed modeling template and hybrid
Gibbs--ESS--Metropolis implementation.  Section \ref{sec:dgpal} extends this
framework to AL through prediction and acquisition. Section
\ref{sec:experiments} provides extensive empirical results on simulated and
real computer experiments, and Section \ref{sec:conclude} concludes with a
discussion.

\section{Review} 
\label{sec:review}

Here we review elements in play in this manuscript, setting the stage for our
contribution: Gaussian process (GP) surrogate modeling, active learning (AL)
or sequential design, and deep GPs (DGPs).

\subsection {Gaussian process surrogates} \label{sec:gp} 

We regard an expensive computer simulation as an evaluation of a blackbox
function $f:\mathbb{R}^d \rightarrow \mathbb{R}$.  To help manage simulation
costs, it is common to develop a surrogate $\hat{f}_n:\mathbb{R}^d \rightarrow
\mathbb{R}$ that approximates $f$ based on example outputs from a small set
of $n$ designed inputs.  Let $X_n$ denote an $n\times d$ training design of
input locations and $Y_n = f(X_n)$ denote the corresponding function
evaluations. The canonical GP surrogate assumes a multivariate normal
distribution (MVN) over the response,
\[
Y_n \sim \mathcal{N}_n\left(\mu, \Sigma(X_n)\right).
\]
To streamline notation, denote $\Sigma_n = \Sigma(X_n)$.  The mean $\mu$ may
be linear in columns of $X_n$, but $\mu = 0$ is often sufficient after
centering \citep{santner2018design,gramacy2020surrogates}.

\begin{figure}[ht!]
\centering
\includegraphics[width=17cm,trim=10 10 0 0]{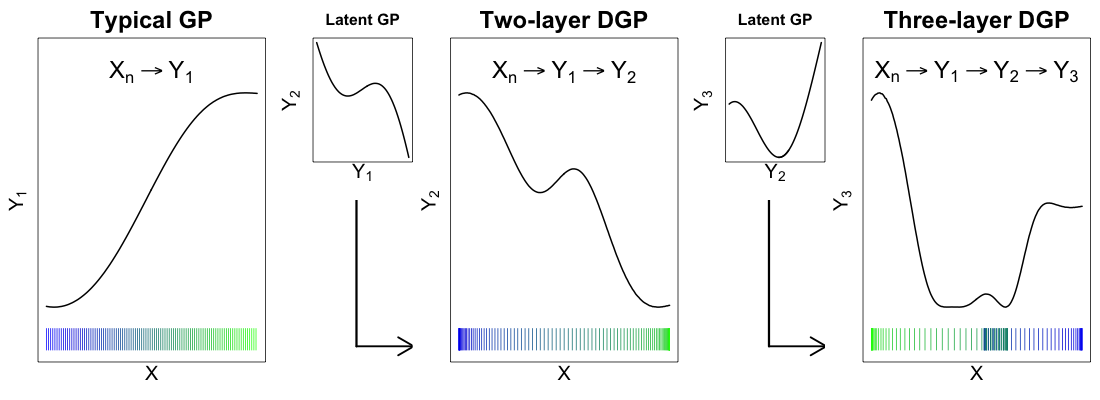}
\caption{GP prior realization passed through latent GPs to generate two- and
	three-layer DGPs.  Colored tick marks on the x-axes show the distribution of
	the latest input ($X_n$, $Y_1$, and $Y_2$). $Y_1$ clusters data at the
	boundaries, and $Y_2$ overlaps data near the middle of the input space.}
\label{fig:dgp}
\end{figure}

The left panel of Figure \ref{fig:dgp} shows a ``prior'' random draw from
such an MVN based on a gridded $X_n$ with $d=1$ under a covariance structure
determined by inverse (squared) Euclidean distance:
\begin{equation}
\label{eq:cov}
\Sigma_n^{ij} = \tau^2 \left(\textrm{exp}\left(-\frac{||x_i-x_j||^2}
	{\theta}\right)+ g\mathbb{I}_{\{i=j\}}\right).
\end{equation}
In this so-called isotropic Gaussian kernel, $\tau^2$, $\theta$, and $g$ act
as hyperparameters controlling the scale, correlation strength (lengthscale),
and noise level (nugget) respectively.  We will often denote $\Sigma_n =
\tau^2
\left(K_\theta(X_n) + g\mathbb{I}_n\right)$ where $K_\theta^{ij} =
\textrm{exp}\left(-||x_i - x_j||^2/\theta\right)$.   Our methods are largely 
agnostic to kernel choice.  For example, a Mat\`{e}rn \citep{stein1999interpolation} 
would swap in nicely.  We prefer an isotropic form in the DGP context, reviewed 
momentarily in Section \ref{sec:deepGP}, for reasons detailed later in Section
\ref{sec:modeling}. Extensions to so-called separable/anisotropic kernels
allowing for lengthscales $\theta_k$ in each input direction, $k=1,\dots,d$,
are preferred in typical GP regression settings.

Given (potentially noisy) training data examples $D_n=(X_n, Y_n)$, 
we would want to learn these hyperparameters. The MVN model structure emits the 
following (marginal) log likelihood,
\begin{equation}
\label{eq:logl}
\log\mathcal{L}\left(Y_n \mid X_n\right) \propto -\frac{1}{2}
	\log |\Sigma_n| - \frac{1}{2}Y_n^\top\Sigma_n^{-1}Y_n.
\end{equation}
Maximum likelihood estimates are commonly used as plug-ins for $\tau^2$, $\theta$, and $g$.
MLE $\hat{\tau}^2$ has a closed form.  Derivative-based numerical maximization
is required for $\hat{\theta}$ and $\hat{g}$.  Details and implementation are
provided by \citet[e.g.,][Chapter 5]{gramacy2020surrogates}.  Here we promote
Bayesian posterior sampling, thinking ahead to DGPs. One can marginalize
$\tau^2$ out of the posterior analytically under a reference prior ($\pi(\tau^2) 
\propto 1/\tau^2$), or any conditionally conjugate inverse Gamma
specification; however, $\theta$ and $g$ require rejection-based MCMC schemes
(i.e., Metropolis--Hastings).  Details are 
provided, e.g., by \citet[][Section 5.5]{gramacy2020surrogates}.  When
training data sizes are low, it can be crucial to average over posterior
uncertainty for $\theta$ and $g$ which, together, facilitate a signal-to-noise
trade-off. 

\begin{figure}[ht!]
\centering
\includegraphics[width=18cm,trim=5 10 0 0]{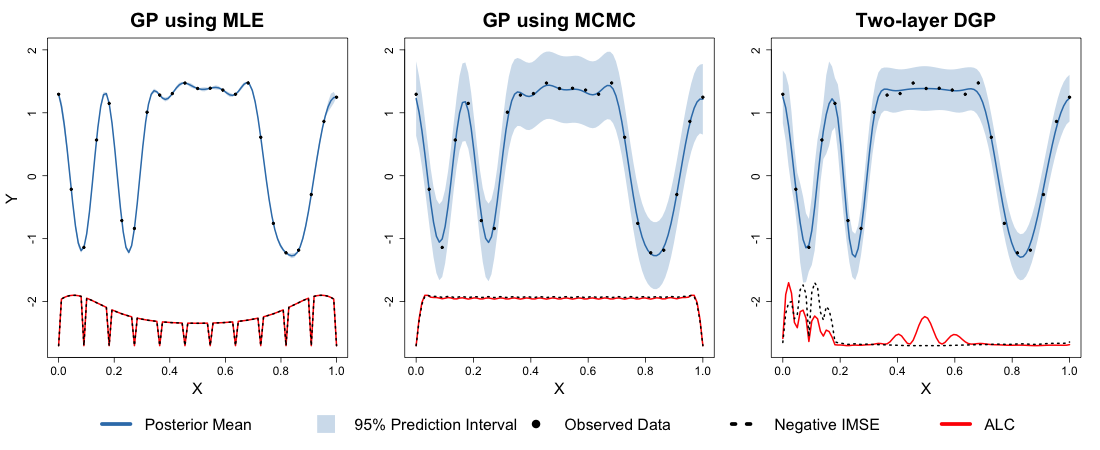}
\caption{Surrogates $\hat{f}_n$ fit to data from  
	$f(x)$ in Eq.~(\ref{eq:oned}): (left) using MLE $\hat{g}$ and 
	$\hat{\theta}$; (center) using MCMC sampling of $g$ and $\theta$; 
	(right) using a two-layer DGP (described in Section \ref{sec:modeling}).  
	ALC and IMSE (negated in order to maximize) are plotted along a 
	dense grid of candidate locations. }
\label{fig:ex}
\end{figure}

Given $D_n$, and under settings of hyperparameters (either MLE or via
posterior sampling),  the posterior predictive distribution for an
$n'\times d$ matrix of testing locations $\mathcal{X}$ has closed form
\begin{equation}
\begin{aligned}
Y(\mathcal{X}) \mid D_n &\sim \mathcal{N}_{n'}\left(\mu_Y(\mathcal{X}),
	\Sigma_Y(\mathcal{X})\right), & \label{eq:pred}
\mbox{where } \quad \mu_Y(\mathcal{X}) &= \Sigma(\mathcal{X}, X_n)\Sigma_n^{-1}Y_n \\
&& \Sigma_Y(\mathcal{X}) &= \Sigma(\mathcal{X}) - \Sigma(\mathcal{X}, X_n)
	\Sigma_n^{-1}\Sigma(X_n, \mathcal{X}),
\end{aligned}
\end{equation}
and $\Sigma(X_n, \mathcal{X})$ is an $n \times n'$ matrix derived by
extending the kernel across training and testing elements.  These equations,
coupled with an inferential strategy (e.g., MLE $\hat{\tau}^2$ converts
Gaussian to Student-$t$), complete the description of surrogate $\hat{f}_n$. 
A 1d illustration is provided in the left and center panels of Figure \ref{fig:ex}, based on MLE
and full posterior sampling, respectively. The GP surrogate's flexibility
provides accurate posterior mean estimates in the non-linear regions, but MLE
point-estimates (left) lead to over-fitting -- interpreting noise as signal
-- in the flat region.  Full posterior sampling (center) appropriately averages
over posterior evidence for both regimes.  

The conditionally analytic structure of GP prediction (\ref{eq:pred}) is
convenient, and its UQ capability -- particularly via MCMC as in the middle
panel of Figure \ref{fig:ex} -- supports a wide range of downstream
applications such as Bayesian optimization
\citep{jones1998efficient,picheny2016bayesian}, computer model calibration to
field data \citep{kennedy2001bayesian,higdon2004combining}, and input
sensitivity analysis \citep{saltelli2002making,oakley2004probabilistic,
marrel2009calculations}.  But GPs are no panacea.  Working with MVNs involves
dense $n \times n$ matrices, requiring quadratic-in-$n$ storage and cubic
decomposition.  Meanwhile, covariance based solely on relative (inverse
Euclidean) distance can be too rigid for dynamics involved in many
simulations, limiting the GP's ability to capture features that change
positionally in the input configuration space.  This so-called stationarity
limitation is coupled, from a practicality perspective, with computational
bottlenecks.  Even if stationarity is relaxed, multiple regimes may not be
recognized without large amounts of training data.

\subsection{Active learning} \label{sec:al}

The most common designs for computer simulation experiments are based on
space-filling principles, either along the margins with Latin hypercube
samples \citep[LHSs;][]{mckay2000comparison}, via pairwise distances
\citep{johnson1990minimax}, or hybrids thereof \citep{morris1995exploratory}. 
These are fine strategies, but they presume a fixed budget of $n$ runs in advance.
Sequential adaptations exist but do not leverage model uncertainties.  For
a review, see \citet[][Chapter 4]{gramacy2020surrogates}. Model-based
analogues, which devise criteria via prior--posterior information gain
\citep[so-called maximum entropy designs;][]{shewry1987maximum} and posterior
mean-squared prediction error (MSE), offer similar behavior.  But these have
a chicken-or-egg problem because they condition on (a priori unknown)
hyperparameter settings. Sequential adaptations, suitably initialized
\citep{zhang2021distance}, offer a natural remedy.  See \citet[][Chapter
6]{gramacy2020surrogates}.

Much recent work on sequential design for GPs comes from ML as a branch of
reinforcement learning called active learning (AL).  We adopt their
terminology of solving acquisition functions (sequential design criteria) to
``actively'' select the data which the model is trained on: begin (0) with a
small training data set $D_n=(X_n, Y_n)$ of size $n=n_0$; then (1) fit a
flexible model to $D_n$; and (2) select $x_{n+1}$ to augment the data $n
\rightarrow n+1$; and (3) repeat from (1).  This setup avoids many pathologies
inherent to static/batch training through a baby-steps approach and allows
progress to be monitored. Maximum-variance acquisitions (for step 2) were
shown to approximate maximum entropy designs without the burden of setting
unknown hyperparameters \citep{mackay1992information}.  Aggregate
variance-based acquisitions \citep{cohn1994neural} approximately minimized
MSE. Both ideas were originally for neural networks (step 1), but were
subsequently ported to GPs \citep{sambu2000gaussian}.  Acquisition based on
expected improvement could target minima in blackbox simulation output
\citep[e.g.,][Chapter
7]{snoek2012practical,picheny2016bayesian,gramacy2020surrogates}.

Here we emphasize building up a simulation campaign where runs are acquired to
reduce MSE. Let $X_{n+1} = \{x_1, \dots, x_n, x_{n+1}\}$ represent the
combined set of $n$ current input locations and new input location $x_{n+1}$.
Let $\sigma^2_n(x)$ denote the posterior predictive variance (i.e., the MSE)
at location $x$ calculated from  $X_n$, i.e., via Eq.~(\ref{eq:pred}) with
singleton $\mathcal{X} = \{x\}$.  Take hyperparameter settings from MLE
calculations or as posterior samples conditioned on data $D_n$.  Now, let
$\breve{\sigma}^2_{n+1}(x)$ denote the deduced posterior predictive variance
at location $x$ calculated from the augmented $X_{n+1}$, but otherwise with
hyperparameter settings based on $D_n$.
\[
\breve{\sigma}_{n+1}^2(x) = \Sigma(x) - \Sigma(x, X_{n+1})\Sigma_{n+1}^{-1}
		\Sigma(X_{n+1}, x) 
\]
We follow the convention of defining both $\sigma_n^2(x)$ and
$\breve{\sigma}_{n+1}^2(x)$ for the purpose of AL via $\Sigma(x) = 1$
rather than $\Sigma(x) = 1 + g$, targeting the variance of the latent random
field (i.e., the variance of the mean) rather than the full predictive
variance.  Making this, and other kernel details explicit ...
\[
\breve{\sigma}_{n+1}^2(x)	= \tau^2\left(1- K_\theta(x, X_{n+1})\left(K_\theta(X_{n+1})
	 + g\mathbb{I}_{n+1}\right)^{-1}K_\theta(X_{n+1}, x)\right).
\]
Hence $\breve{\sigma}_{n+1}^2(x) \rightarrow 0$ as $n\rightarrow \infty$,
providing a natural ``asymptote for learning.''
Observe that the nugget $g$ is still involved in $\Sigma_{n+1}^{-1}$.

It is sensible to acquire new $x^\star_{n+1}$ to minimize
$\breve{\sigma}_{n+1}^2(x)$ integrated over all $x$, 
\[
x_{n+1}^\star =
\operatorname*{argmin}_{x_{n+1}}\; \textrm{IMSE}(X_{n+1}) \quad\textrm{where}\quad
\textrm{IMSE}(X_{n+1}) = 
\int \breve{\sigma}_{n+1}^2(x) \, dx.
\]
When the input space used in the domain of integration is a hyper-rectangle, a
closed form expression is available.  See, e.g., \citet{binois2019replication}
for coded inputs $[0,1]^d$ and extensions by \citet{cole2021locally} to $[a,
b]^d$ and Gaussian measures over inputs.  

Despite a degree of analytic tractability, calculating IMSE is cumbersome,
especially in the Bayesian posterior sampling context.  Varying
hyperparameterization thwarts many pre-calculations and matrix decomposition
tricks that would otherwise lend a degree of efficiency.  Instead we prefer an
approximation obtained by extending Monte Carlo integration over posterior
draws to include the input space.  This is the acquisition criteria now known
as ``active learning Cohn'' \citep[ALC;][]{cohn1994neural,sambu2000gaussian}.
Observe that minimizing IMSE over $x_{n+1}$ is equivalent to maximizing the
difference between $\sigma^2_n(x)$ and $\breve{\sigma}^2_{n+1}(x)$.
\begin{align*}
\Delta\sigma_n^2(X_{n+1}) &= \int \sigma^2_n(x) - \breve{\sigma}_{n+1}^2(x) dx 
	\propto \int \tau^2 K_\theta(x, X_{n+1})\left(K_\theta(X_{n+1}) + 
		g\mathbb{I}_n\right)^{-1}K_\theta(X_{n+1}, x) dx
\end{align*}
The ALC criterion approximates this integral with a sum over a reference set,
$X_{\mathrm{ref}}$, 
\begin{equation}
\label{eq:alc}
\begin{aligned}
\textrm{ALC}(X_{n+1} \mid X_{\mathrm{ref}}) &\propto \sum_{x\in X_{\mathrm{ref}}} 
	\tau^2 K_\theta(x, X_{n+1}) \left(K_\theta(X_{n+1}) + g\mathbb{I}_{n+1}\right)^{-1}
	K_\theta(X_{n+1}, x) \\
\end{aligned}
\end{equation}
with acquisition as $x_{n+1}^\star = \mathrm{argmax}_{x_{n+1}}
\textrm{ALC}(X_{n+1}
\mid X_{\mathrm{ref}})$. A uniform reference 
set yields proportional ALC and IMSE surfaces.  See left and center panels of
Figure \ref{fig:ex}.  Instead of library-based optimization \citep{gramacy2015local},  
we prefer discrete search over (possibly random)
candidates because this lends well to averaging across MCMC
iterations and parallel implementation. 

Even with variance-based criteria, AL is
limited by the stationarity of the covariance kernel of the GP
\citep{gramacy2009adaptive}.  In Figure \ref{fig:ex}, the MLE surface
(left) gets the ``wrong answer'' as regards AL, discouraging acquisition 
where it needs it most: in the middle of inputs where it has confused noise as
signal.  With MCMC (center), the ALC/IMSE criteria indentify a more uniform 
preference for the next run, although both ultimately prefer runs near the edges,
which manifests as a form of leverage in this case.  Neither MLE nor MCMC-based fits are able to
recognize what human intuition would likely suggest: more data is needed where
the function is wiggly than where it is not.  This is not the fault of the AL
criteria.  The stationary GP is not ``allowed'' to recognize two (or more)
distinct regimes, of wiggly and flat.  
Only after relaxing stationarity, so that predictive variance is not simply a
matter of distance to nearby training values, can we obtain a notion of
predictive uncertainty which is region-dependent.  

\subsection{Deep Gaussian processes} \label{sec:deepGP} 

A DGP is a hierarchical layering of GPs where each layer is conditionally MVN.
This definition is inherently broad and lends to various formulations.
\citet{dunlop2018how} detailed four methods for constructing DGPs, each
providing different levels of feasibility and interpretability. We view DGPs
as functional compositions, arguably the most interpretable and easily
implemented option.

In a DGP prior, the inputs $X_n$ are mapped through one or more intermediate
GPs before reaching the response $Y_n$. The inputs to those intermediate
``layers'' act as latent variables, warping the input space but remaining
unobserved.  Henceforth, we shall refer to typical GP regression as described
in Section \ref{sec:gp} as a ``one-layer'' GP.  A two-layer model, with inputs
to the new hidden GP labeled as $W$, may be described hierarchically as
\begin{equation}
\label{eq:dgp}
\begin{aligned}
Y_n \mid W &\sim \mathcal{N}_n\left(0, \Sigma(W)\right) &\quad \quad
W &\sim \mathcal{N}_n\left(0, \Sigma(X_n)\right).
\end{aligned}
\end{equation}
Conditional on $\Sigma$, the marginal likelihood 
may be represented as an integral over unknown $W$,
\begin{equation}
\label{eq:int}
\mathcal{L}(Y_n \mid X_n) \propto \int \mathcal{L}(Y_n \mid W)\mathcal{L}(W\mid X_n) \;dW
\end{equation}
where $\log\mathcal{L}(Y_n\mid W)$ and $\log\mathcal{L}(W\mid X_n)$ are
defined as in Eq.~(\ref{eq:logl}).  A three-layer model will have two hidden
layers (denoted $Z$ and $W$) and will involve three MVN likelihoods with a 
double integral over $Z$ and $W$.  The center and right panels of Figure
\ref{fig:dgp} show the hierarchical composition of two- and three-layer DGPs in
a single dimension, accumulating from the left panel's ordinary, one-layer GP.
As uniformly distributed $X_n$ are fed through intermediate layers they are
re-distributed, and response values are no longer stationary.  Intermediate
layers are not confined to a single dimension. In fact, a single latent
dimension is often inadequate. We will refer to multiple component dimensions
of intermediate layers as ``nodes'', following deep learning nomenclature.

Adding additional depth and nodes to intermediate layers will eventually yield
diminishing returns.  \citet{damianou2013deep} found some justification for a
five-layer DGP in classification tasks, but two- and three-layer DGPs have been
sufficient for real-valued outputs common to computer surrogate modeling 
\citep{radaideh2020surrogate}. \citet{dunlop2018how} similarly preferred
two and three layers.  We find that low numbers of latent nodes (no higher than
the dimension of the input space) and limited depth (no deeper than three
layers) are sufficient for surrogate modeling and AL.  More
details are provided in Section \ref{sec:modeling} with evidence in Section
\ref{sec:experiments}.  The benefit of DGPs from the perspective of AL is
foreshadowed in the right panel of Figure \ref{fig:ex}. A
degree of nonstationarity, as facilitated by warping the input space so that
some pairs are (effectively) closer than others when calculating covariance,
is sufficient to nudge acquisitions towards the ``interesting'' part of the
input space,{} rather than being essentially space-filling.

Closed form Bayesian inference for two- and three-layer DGPs is intractable
since it is not possible to analytically integrate the hidden layer(s) out of
the (marginal) likelihood (\ref{eq:int}).  
Variational inference and other maximization--oriented methods
\citep{damianou2013deep,salimbeni2017doubly} are computationally thrifty, but yield
probabilistic representations which can over-simplify, particularly as regards
UQ.  MCMC methods address uncertainty more thoroughly at the expense of computation.  
To mitigate these costs, efficient mixing of the Markov chain is essential.   

\section{Modeling and inference} 
\label{sec:modeling}

Here we outline a novel elliptical slice sampling-based
scheme for posterior inference with DGPs.

\subsection{Model specification}\label{sec:modelspec}

In a DGP, extreme
flexibility can come at the cost of identifiability and practicality. Here we
provide a template that has worked well for surrogate modeling in several
realistic settings, as showcased in Section \ref{sec:experiments}, and which
supports AL efforts (Section \ref{sec:dgpal}). We begin with a
two-layer setup (\ref{eq:dgp}), defining the hierarchical structure in terms
of distance-based covariance (\ref{eq:cov}).
\begin{equation}\label{eq:twolayer}
\begin{aligned}
Y_n \mid W &\sim \mathcal{N}_n\left(0, \tau^2(K_{\theta_y}(W) + 
	g\mathbb{I}_n)\right) & \quad\quad
	   W &\sim \mathcal{N}_n\left(0, K_{\theta_w}(X_n)\right)
\end{aligned}
\end{equation}
Notice that scale $\tau^2$ and nugget $g$ are placed on the outer layer only.
Specifying noiseless hidden layers with unit scale is crucial, we find, to
stable posterior inference.

Although the notation in Eq.~(\ref{eq:twolayer}) is suggestive of a
one-dimensional $W$, in practice it can be beneficial to have multiple nodes,
i.e., a multi-dimensional $W$.  We find that $\mathrm{dim}(W)
=\mathrm{dim(X_n)} = d$ works well as a default, although smaller values
may be appropriate for high-dimensional $X_n$. Let $W_j$ represent the
$j^\mathrm{th}$ node of $W$ for $j = 1, \dots, p$.  Note that, like a
column of $X_n$ or response $Y_n$, $W_i$ is an $n$-vector with one
coordinate for each training data point. Although nodes of $W$ could be
modeled jointly with cross covariances, we recommend the following
simplifications.

\begin{itemize}
\item[i.] $W_j$ and $W_k$, for $j \ne k$, are conditionally independet \citep{damianou2013deep}.
\item[ii.] Each $W_j$ is modeled as isotropic (\ref{eq:cov}) in inputs $X_n$
via unique scalar lengthscale $\theta_w[j]$, for $j=1,\dots,p$, regardless
of the input dimension $d$.
\item[iii.] Similarly, $Y_n$ is modeled as isotropic in all nodes of $W$ with scalar 
lengthscale $\theta_y$.
\end{itemize}
\begin{figure}[ht!]
\centering
\includegraphics[width=9cm,trim=0 20 0 30]{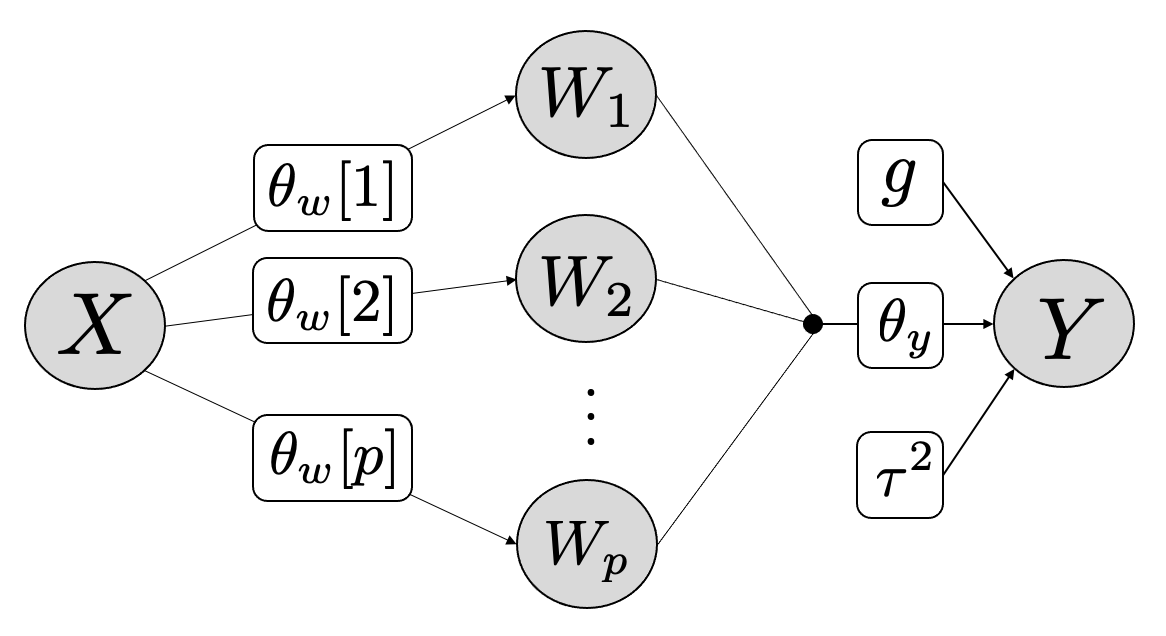}
\caption{Model structure for a two-layer DGP with $p$ latent nodes.}
\label{fig:twolayer}
\end{figure}

Figure \ref{fig:twolayer} depicts this model structure, still focusing on the
two-layer setup. Impositions ii.--iii.~are essential to reign in complexity.
Latent $W$, even with a single node but especially for $p \gg 1$, offer more
than enough flexibility to ``imitate'' a separable/anisotropic kernel
without the added complexity of additional lengthscale parameters.  This is
simplest to see with $p = d$.  If the data desire lower correlation in one
coordinate direction compared to another, then components of $W$ can organize
themselves so that $W_j$, say, ``fans out'' more than $W_k$ does.  Of course, it is
easy to imagine more complicated arrangements.  The filled circle
between the $W$ and $Y$ layers in the diagram represents imposition iii.,
depicting that all nodes of $W$ form inputs to the outer layer for $Y$ under
isotropic (scalar) lengthscale $\theta_y$.

Building off that setup, consider a three-layer model built as follows.
\begin{align*}
Y_n \mid W &\sim \mathcal{N}_n\left(0, \tau^2(K_{\theta_y}(W) + g\mathbb{I}_n)\right) & \quad
 W \mid Z &\sim \mathcal{N}_n\left(0, K_{\theta_w}(Z)\right) \\
&&	   Z &\sim \mathcal{N}_n\left(0, K_{\theta_z}(X_n)\right)
\end{align*}
Here, as above, conditional independence and isotropy (i.-ii.)~for $W$ is duplicated
in the new latent layer $Z$.  We have found it helpful to restrict the number of 
nodes in $Z$ to match $W$, i.e., $\mathrm{dim}(Z) = \mathrm{dim}(W) = p \leq d$.
Again, each $Z_j$ is an $n$-vector, for $j=1,\dots,n$, of latent variables.
\begin{figure}[ht!]
\centering
\includegraphics[width=11cm,trim=0 20 0 30]{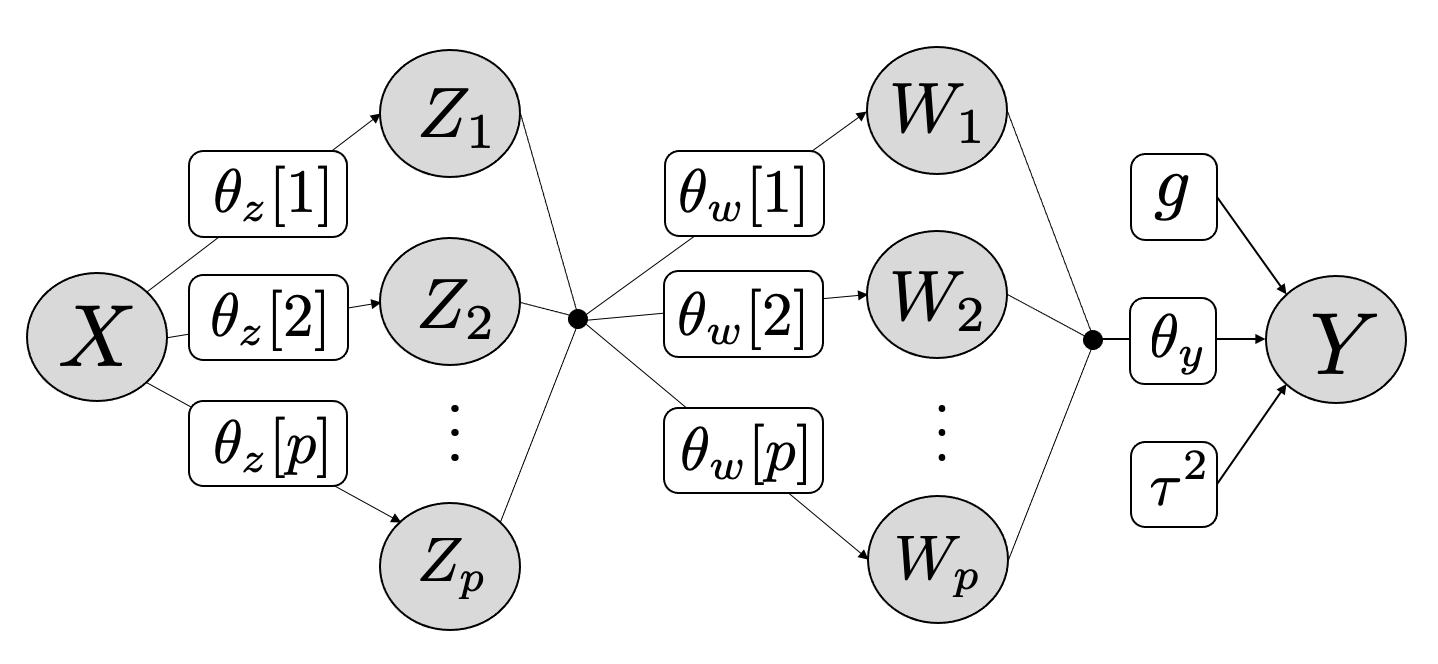}
\caption{Model structure for a three-layer DGP with $p$ latent nodes in $Z$
and $W$.}
\label{fig:threelayer}
\end{figure}
Figure \ref{fig:threelayer} depicts this structure diagrammatically. 
The filled circle between $Z$ and $W$ layers,
with edges fanning out in both directions, is intended to represent
communication between all $Z$ and $W$ nodes, like an ``information bus''
capturing $p^2$ interconnections.  Each latent $Z_j$ input, $j=1,\dots,p$, is
connected to each $W_k$ output through an inverse-distance-based covariance
kernel with (scalar) lengthscale parameters $\theta_w[k]$, for $k=1,\dots,p$.

\subsubsection*{Hyperparameter priors}

We complete our Bayesian hierarchical model specification with the following
priors on hyperparameters, comprising lengthscales $\theta = \{\theta_y,
\theta_w, \theta_z\}$, scale $\tau^2$, and nugget $g$ on the outer layer.
Conditionally conjugate inverse Gamma (IG) priors are common for scales
like $\tau^2$ because they may
subsequently be analytically integrated out of an otherwise numerical
posterior sampling scheme.  Details are provided in Section \ref{sec:sampling}
shortly.  Specifying IG parameters of zero \citep[following the description
in][]{gelman2013bayesian} leads to a so-called reference prior $\pi(\tau^2)
\propto 1/\tau^2$ \citep{berger2001objective}, which is a common default in GP
spatial modeling. Although improper, the posterior is proper (in our zero-mean
context) as long as $n \geq 1$, i.e., as long as there is at least one
training data example.

We prefer proper priors for the other parameters, and choose members of the
Gamma family as their familiar form offers convenient specification, which we
tailor to penalize against pathological settings such as $\theta \rightarrow
\{0,\infty\}$.  Specifically, consider $\{\theta,g\}
\stackrel{\mathrm{iid}}{\sim} \mathrm{G}(3/2, b_{[\cdot]})$ where
$b_{[\cdot]}$ is adjusted depending on the hyperparameter in
question. This choice mirrors other work in Bayesian posterior sampling of GP
hyperparameters \citep[e.g.,][]{gramacy2008bayesian} and in penalized marginal
maximum likelihood settings \citep[e.g.,][]{gramacy2015local}. We find it
helpful to nudge the posterior toward a hierarchy in the latent nodes with
$b_{[\theta_y]} > b_{[\theta_w]} > b_{[\theta_z]}$ encoding a prior belief 
that as layers get deeper they should be less ``wiggly'' (closer to the identity 
mapping). Particular, user-adjustable, default values for $b_{[\cdot]}$ provided 
in our software are detailed in Section \ref{sec:implement}.  When modeling a deterministic
computer model simulation in the outer layer we fix $g = \varepsilon$, a small
non-zero value such as \verb!sqrt(.Machine$double.eps)! in {\sf R}, to
interpolate those runs.

\subsection{MCMC posterior sampling} \label{sec:sampling}

Here we propose a hybrid Gibbs--ESS--Metropolis scheme for sampling
hyperparameters and latent layers.  A closed form GP marginal likelihood,
integrating out scale $\tau^2$ under the reference prior, is crucial to the
subsequent development.  Although the calculation is rather textbook, e.g., see
\citet[][Section 5.5, exercise 1]{gramacy2020surrogates}, we provide it
below in our notation for concreteness.
\begin{equation}
\label{eq:marglikouter}
\log \mathcal{L}(Y_n \mid W, \theta_y, g) \propto -\frac{n}{2}
	\log\left(n \hat{\tau}^2 \right)
	- \frac{1}{2}\log |K_{\theta_y}(W) + g\mathbb{I}_n|
	\; \mbox{ with } \; 
	\hat{\tau}^2 = \frac{Y^\top(K_{\theta_y}(W) + g\mathbb{I}_n)^{-1}Y}{n}
\end{equation}
Throughout, relationship ``$\propto$'' for log likelihoods denotes
that an additive constant has been dropped. Taking $W \equiv X_n$ lends explicit
form to Eq.~(\ref{eq:logl}) for the ordinary GP case, where $\hat{\tau}^2$ may
be interpreted as an MLE. 
For DGPs with two and three layers, we additionally need
\begin{equation}
\label{eq:marglik2}
\begin{aligned}
\log \mathcal{L}(W \mid Z, \theta_w) & \propto \sum_{i=1}^p \log\mathcal{L}
	\left(W_i \mid Z, \theta_w[i] \right)\\
	&\propto \sum_{i = 1}^p\left(-\frac{1}{2}
	\log |K_{\theta_w[i]}(Z)|  - \frac{1}{2}W_i^\top
		\left(K_{\theta_w[i]}(Z)\right)^{-1} W_i \right).
\end{aligned}
\end{equation}
In two-layers, take $Z \equiv X_n$ and collect 
\[
\log \mathcal{L}(Y_n \mid W, X_n, \theta, g) = \log \mathcal{L}(Y_n \mid W, \theta_y, g) + 
\log \mathcal{L}(W \mid X_n, \theta_w),
\]
combining Eqs.~(\ref{eq:marglikouter}--\ref{eq:marglik2}).  Of course, in the
context above of unknown (latent) $W$, the quantity $\log \mathcal{L}(W \mid
X_n, \theta_w)$ is actually a log prior, but its form is
nevertheless given by Eq.~(\ref{eq:marglik2}), so we find it convenient to
notate using marginal log likelihoods.

Finally, in the three layer case (with trivial extension beyond), we have
\[
\log \mathcal{L}(Y_n \mid W, Z, X_n, \theta, g) = \log\mathcal{L}(Y_n \mid W, \theta_y, g) + 
\log \mathcal{L}(W \mid Z, \theta_w) + \log \mathcal{L}(Z \mid X_n, \theta_z)
\]
where the third term (and beyond) follows Eq.~(\ref{eq:marglik2}) for $Z$
and $X_n$ instead of $W$ and $Z$. 
The posterior distribution is completed with the hyperparameter
priors outlined at the end of Section \ref{sec:modelspec},
\[
\pi(W, Z, \theta, g \mid D_n) \propto \mathcal{L}(Y_n \mid W, Z, X_n, \theta, g)
\times \pi(\theta, g),
\]
and the remainder of this section details how we obtain samples in the Gibbs
framework summarized in Algorithm \ref{alg:gibbs}.  Hyperparameters
$\theta$ and $g$ follow a somewhat conventional random-walk
Metropolis--Hastings (MH) procedure.  To acknowledge a restricted support
$[\varepsilon,\infty]$, proposals $(q)$ follow a ``uniform sliding window'' scheme first 
outlined by \citet{gramacy2008bayesian} whereby
\begin{equation}\label{eq:proposal}
\theta^\star \sim \textrm{Unif}\left(\frac{\ell\theta^{(t-1)}}{u}, 
	\frac{u\theta^{(t-1)}}{\ell}\right) 
	\quad  \mbox{ so that } \quad \frac{q(\theta^{(t-1)} \mid \theta^\star)}
	{q(\theta^\star \mid \theta^{(t-1)})} = \frac{\theta^{(t-1)}}{\theta^\star}
\end{equation}
 features as the proposal ratio in the MH acceptance probability.  We sample
all components of $\theta$ and $g$ (except when $g\rightarrow \varepsilon$ for
interpolating deterministic simulations) in this way.  Defaults for
tuning parameters $u$ and $\ell$ are given in Section \ref{sec:implement}. Under
our template, each hyperparameter is involved in only one likelihood
component. Thus each MH acceptance ratio requires only one
MVN density evaluation. For example, acceptance for $\theta_y^\star$ is based on
\begin{equation}
\label{eq:mhalpha}
\alpha = \min \left(1, \frac{\mathcal{L}\left(Y_n\mid W, \theta_y^\star, g\right)
	\pi\left(\theta_y^\star\right)}
	{\mathcal{L}\left(Y_n\mid W, \theta_y^{(t-1)}, g\right)\pi\left(\theta_y^{(t-1)}\right)} 
	\times \frac{\theta_y^{(t-1)}}{\theta_y^\star}\right).
\end{equation}
The likelihood component involved in each Metropolis--within-Gibbs step is
detailed in Algorithm~\ref{alg:gibbs}.  To shorten this algorithm for a
two-layer model, simply remove sampling of $\theta_z$ and $Z$ and replace
remaining instances of $Z$ with $X_n$.

\medskip
\begin{algorithm}[H] 		
\DontPrintSemicolon
initialize $g^{(1)}$, $\theta_y^{(1)}, \theta_w^{(1)}$, $\theta_z^{(1)}$, $W^{(1)}, 
$and $Z^{(1)}$\;
\For{$t = 2, \dots, T$}{
	$g^{(t)} \sim \pi(g\mid Y_n, W^{(t-1)}, \theta_y^{(t-1)}, g^{(t-1)})$ 
	\tcp*{MH (\ref{eq:mhalpha}) via $\mathcal{L}(Y_n\mid W, \theta_y\ g)$}
	$\theta_y^{(t)} \sim \pi(\theta_y\mid Y_n, W^{(t-1)}, \theta_y^{(t-1)}, g^{(t)})$ 
	\tcp*{MH (\ref{eq:mhalpha}) via $\mathcal{L}(Y_n\mid W, \theta_y\ g)$}
	\For{$i = 1, \dots, p$}{
	$\theta_w[i]^{(t)} \sim \pi(\theta_w[i]\mid W_i^{(t-1)}, Z^{(t-1)}, \theta_w[i]^{(t-1)})$ 
	\tcp*{MH (\ref{eq:mhalpha}) via $\mathcal{L}(W_i\mid Z, \theta_w[i])$}
	}
	\For{$i = 1, \dots, p$}{
	$\theta_z[i]^{(t)} \sim \pi(\theta_z[i]\mid Z_i^{(t-1)}, X_n, \theta_z[i]^{(t-1)})$
	\tcp*{MH (\ref{eq:mhalpha}) via $\mathcal{L}(Z_i\mid X_n, \theta_z[i])$}
	}
	\For{$i = 1, \dots, p$}{
	$W_i^{(t)} \sim \pi(W \mid Y_n, W_{\geq i}^{(t-1)}, W_{< i}^{(t)}, Z^{(t-1)},
		g^{(t)}, \theta_y^{(t)}, \theta_w^{(t)})$
		\tcp*{ESS via (\ref{eq:walpha})}
	}
	\For{$i = 1, \dots, p$}{
	$Z_i^{(t)} \sim \pi(Z \mid X_n, W^{(t)}, Z_{\geq i}^{(t-1)}, Z_{< i}^{(t)}, 
		\theta_w^{(t)}, \theta_z^{(t)})$ 
		\tcp*{ESS via (\ref{eq:zalpha})}
	}
}
\caption{Gibbs sampling procedure for three-layer DGP}
\label{alg:gibbs}
\end{algorithm}
\medskip

Latent $W$ and $Z$ could be handled similarly, i.e., by MH.  While there 
are a number of approaches for sampling Gaussian fields in non-conjugate 
settings \citep[e.g.,][]{neal2011mcmc,oliver1997markov,higdon2003markov},
such rejection-based samplers may mix poorly even with tedious
fine tuning of proposal functions.  We think this is why MCMC 
inference for DGPs is often dismissed in preference for maximization--based shortcuts.
However, we find elliptical slice sampling \citep[ESS;][]{murray2010elliptical}
of $W$ and $Z$ to be particularly efficient.

To adapt ESS for DGP latent layers, allow us to briefly digress to the general
case in which we desire samples from a posterior under a zero-mean MVN prior, 
$\pi(f) \propto \mathcal{L}(f)\times\mathcal{N} (f; 0, \Sigma)$. ESS initializes with
a random draw from the prior $f^\mathrm{prior} \sim \mathcal{N}(0, \Sigma)$ and
a random angle $\gamma \sim
\mathrm{Unif}(0, 2\pi)$ with consequent boundaries set to $\gamma_{\min} =
\gamma - 2\pi$ and $\gamma_{\max} = \gamma$.  Proposal $f^\star$ is a function
of the previous iteration, the prior draw, and the angle,
\begin{equation}\label{eq:ess}
f^\star = f^{(t-1)}\cos\gamma + f^\mathrm{prior}\sin\gamma
\quad \mbox{with acceptance probability~} \quad
\alpha = \min\left(1, \frac{\mathcal{L}(f^\star)}{\mathcal{L}(f^{(t-1)})}\right).
\end{equation}
Upon rejection, ESS shrinks the ``bracket'' on the angle -- specifically, setting
$\gamma_{\min} = \gamma$ if $\gamma < 0$ and $\gamma_{\max} = \gamma$
otherwise -- and re-samples $\gamma \sim \mathrm{Unif}(\gamma_{\min},
\gamma_{\max})$, yielding new $f^\star$. The process is repeated, shrinking
the bracket and re-proposing, until acceptance, constituting one iteration/one
sample.  This distinguishes ESS from MH methods in which
rejections result in identically repeated values and poor mixing.  Although
ESS may require more likelihood evaluations (each new proposal warrants one)
than MH, it tends to yield more ``effective samples'' \citep{kass1998markov}
through enhanced mixing, as demonstrated in Section \ref{sec:dgpillus}, all
without having to set any tuning parameters.

To implement ESS for $W$ and $Z$, one may leverage the likelihood
(\ref{eq:marglik2}) as a prior over the unknown function.  First generate from
the prior for each node,
\[
W_i^\mathrm{prior} \sim \mathcal{N}_n\left(0, K_{\theta_w[i]}(Z)\right)
\quad \mbox{and} \quad 
Z_i^\mathrm{prior} \sim \mathcal{N}_n\left(0, K_{\theta_z[i]}(X_n)\right)\!.
\]
In the two-layer case, set $Z \equiv X_n$ for each $W_i$.  ESS proposals
$W_i^\star$ and $Z_i^\star$ follow Eq.~(\ref{eq:ess}) based on previous
iteration $W_i^{(t-1)}$ and $Z_i^{(t-1)}$ respectively.  Acceptance
probabilities are based on a likelihood component designed to ``receive'' that
latent quantity as an input.  Even though nodes of a layer are independently
proposed, all are involved in the model likelihood (under imposition iii.)~and
hence impact acceptance.  For $W_i$, in either a two- or three-layer
model, that involves Eq.~(\ref{eq:marglikouter}),
\begin{equation}
\label{eq:walpha}
\alpha = \min\left(1, \frac{\mathcal{L}(Y_n \mid W_i^\star, W_{> i}^{(t-1)}, 
	W_{<i}^{(t)}, \theta_y, g)}
	{\mathcal{L}(Y_n \mid W_{\geq i}^{(t-1)}, W_{<i}^{(t)}, \theta_y, g)}\right)
\end{equation}
where, in true Gibbs fashion, we condition on the most recent iteration of each node.
Acceptance probabilities for $Z_i$ use the likelihood of Eq.~(\ref{eq:marglik2}).
\begin{equation}
\label{eq:zalpha}
\alpha = \min\left(1, \frac{\mathcal{L}(W \mid Z_i^\star, Z_{> i}^{(t-1)}, 
	Z_{<i}^{(t)}, \theta_w)}
	{\mathcal{L}(W \mid Z_{\geq i}^{(t-1)}, Z_{<i}^{(t)}, \theta_w)}\right)
\end{equation}
In contrast to $\theta_w[i]$, which only features in one component of this
likelihood, acceptance of node $Z_i$ requires full evaluation of this
likelihood (including all $p$ elements of the sum).  Specific implementation
details, including default priors and proposals, and demonstration of the
mixing of ESS samples of latent layers are provided in Sections
\ref{sec:implement} and \ref{sec:dgpillus}, respectively.

\subsection{Illustration}\label{sec:dgpillus}

MCMC sampling of latent layers $W$ and $Z$ is crucial to downstream
tasks, such as AL in Section \ref{sec:dgpal}.  Their posteriors are inherently
multi-modal, since only pairwise distances feature in the next layer.  Such
lack of identifiability poses a challenge to MCMC mixing and posterior
exploration, but on the flip side of that coin is a convenient barometer for
judging a proposed sampling scheme.
\begin{figure}[ht!]
\centering
\includegraphics[width=15cm,trim=0 20 0 10]{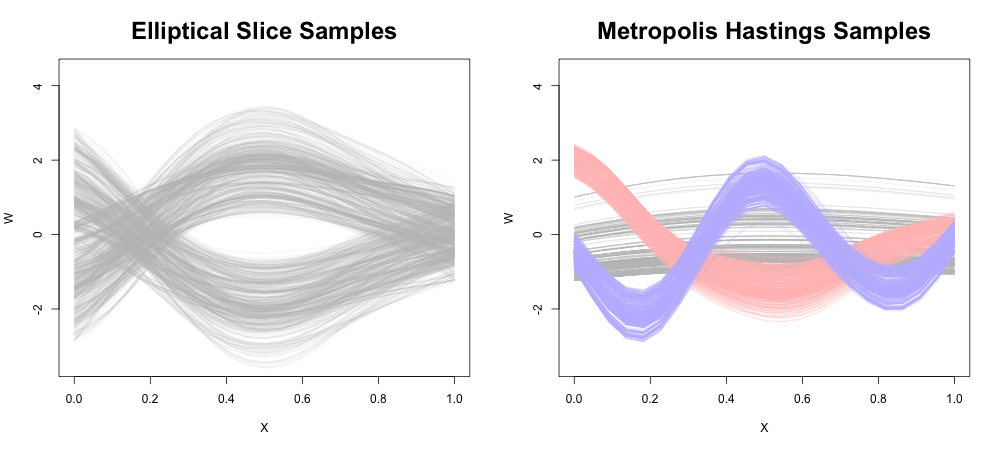}
\caption{ESS and MH samples of latent $W$ from the two-layer DGP of 
Figure \ref{fig:ex}, otherwise following Algorithm
\ref{alg:gibbs} for hyperparameters $\theta$ and $g$. MH samples show three
different initializations; each (three MH \& one ESS) comprise 
10,000 total and 6,000 burn-in iterations with thinning by half.}
\label{fig:ess}
\end{figure}
Our ESS proposal mechanism (\ref{eq:ess}) is particularly well-suited for this
as it can ``bounce'' back and forth between modes.  To demonstrate, consider
the setup in Figure \ref{fig:ex} with a two-layer DGP.  The fitted surface is
shown in the right panel, but we have not yet discussed prediction (Section
\ref{sec:pred}).  For now, focus on latent $W$ via ESS shown in the left panel
of Figure \ref{fig:ess}.  Notice how $W$ samples bend inputs $X$ so that the
middle inputs are distinguished from extremities, separating flat interior
dynamics from outer wiggly ones either by
bending down then back up or up then back down.

To offer contrast, we implemented an MH alternative for sampling $W$ using
kernel-based basis proposals \citep[similar to][]{sejdinovic2014kernel,higdon2003markov} 
conditioned on
sampled $\theta_w^{(t)}$. See the right panel in Figure \ref{fig:ess}.  MH
samples were highly sensitive to initialization and step size.  Depending on
how the chain is initialized, with three random settings shown (and indicated
with color), we get very different $W$, a telltale sign of poor
mixing/convergence of the chain.  In each case shown, the step size was finely
tuned to achieve an acceptance rate of 30-35\%.  This tuning and
re-initialization requires many more MCMC iterations than the ESS comparator
and makes burn-in of the chain difficult to assess.  Posterior predictions
calculated from these MH samples (following Section \ref{sec:pred}, but not
shown) all resulted in poorer out-of-sample root mean square errors (RMSEs)
than those fit via ESS under similar computational budgets.

\section{Active learning}
\label{sec:dgpal}

Given training data $D_n = \{X_n, Y_n\}$, AL se{}eks to optimize a
criterion conceived to assess the potential value of $x_{n+1} \mid D_n$ at
which to acquire a new $y_{n+1} = f(x_{n+1})$, ultimately forming augmented
data $D_{n+1} = \{ X_{n+1}, Y_{n+1} \}$ with $x_{n+1}^\top$ and $y_{n+1}$ in
the last row and entry, respectively. Let $\mathcal{X}$ denote a set of
candidates from which $x_{n+1}$ will be selected.  Criterion in hand,
AL acquisition with DGPs may proceed as follows: (1) collect MCMC samples
indexed by $t\in\mathcal{T}$ (Section
\ref{sec:sampling}), (2) map $\mathcal{X}$ to the criterion for each
$t$; (3) average over $t\in\mathcal{T}$; (4)
choose from $\mathcal{X}$ optimizing the average.

Most AL criteria (e.g., in Section \ref{sec:al} for ordinary GPs) target
predictive quantities, particularly variance.  Consequently we have delayed a
discussion of prediction for DGPs, which could have featured in Section
\ref{sec:modeling}, to here (Section \ref{sec:pred}) in the context of AL
acquisition.  The relevant calculations boil down to propagating
testing inputs $\mathcal{X}$ (i.e., the AL candidates) through latent layers
$\mathcal{X} \rightarrow \mathcal{Z} \rightarrow \mathcal{W} \rightarrow
\mathcal{Y}$ over MCMC iterations $t \in \mathcal{T}$.  Evaluating AL acquisition 
criteria (Section \ref{sec:acq}) amounts to
manipulating those posterior samples, and their generating
mechanism, downstream.

\subsection{Prediction} 
\label{sec:pred}

Here we extend MCMC sampling (Section \ref{sec:sampling}) to posterior
predictive draws $Y(\mathcal{X}) \mid D_n$ at $n' \times d$ testing locations
specified row-wise in $\mathcal{X}$.  For simplicity we limit our discussion
here, and in Section \ref{sec:acq}, to the two-layer DGP case. Extending to
three layers is straightforward, albeit cumbersome notationally.
Implementation details and comparisons are provided for both cases in Section
\ref{sec:experiments}.

After conducting MCMC sampling and any thinning or burn-in (details in Section
\ref{sec:implement}), we are left with samples $\{\hat{\tau}^{2(t)}, g^{(t)},
\theta_w^{(t)}, \theta_y^{(t)}, W^{(t)} \mid t\in\mathcal{T}\}$ from the
posterior distribution of all unknowns.  Although all of these are involved,
development of the relevant predictive quantities hinges most crucially on
$W$. Recall that $W$ follows a GP in $X_n$, and $Y_n$ a GP in $W$. Samples
$W^{(t)}$ in hand, we may utilize predictive equations
(\ref{eq:pred}) to calculate posterior moments over $\mathcal{X}$.  Combining
with Eq.~(\ref{eq:twolayer}) yields that $\mathcal{W}
\equiv W(\mathcal{X})$ is MVN.  For a particular sample indexed by $t$, the
moments are:
\begin{equation}
\label{eq:wmoments}
\begin{aligned}
\mu^{(t)}_{W_i}(\mathcal{X}) &= K_{\theta_w^{(t)}[i]}(\mathcal{X}, X)
	(K_{\theta_w^{(t)}[i]}(X))^{-1} W^{(t)}_i \\
\Sigma^{(t)}_{W_i}(\mathcal{X}) &= K_{\theta_w^{(t)}[i]}(\mathcal{X}) - 
	K_{\theta_w^{(t)}[i]}(\mathcal{X}, X)(K_{\theta_w^{(t)}[i]}(X))^{-1}
	K_{\theta_w^{(t)}[i]}(X, \mathcal{X}).
\end{aligned}
\end{equation}

Sampled $\mathcal{W}^{(t)}$, whose values represent a warping of predictive
locations $\mathcal{X}$, may then be fed into the outer-layer to derive
posterior moments for  $\mathcal{Y} \equiv Y(\mathcal{X})$. Together with
$\hat{\tau}^{2(t)}$ following Eq.~(\ref{eq:marglikouter}) evaluated with
$W^{(t)}$ and $\theta_w^{(t)}$, similar arguments as for $\mathcal{W}$ yield
that  $\mathcal{Y}^{(t)}$ follows a multivariate Student-$t$ predictive
distribution with $n$ degrees of freedom and
\begin{align}
\mu^{(t)}_Y(\mathcal{W}^{(t)}) &= K_{\theta_y^{(t)}}(\mathcal{W}^{(t)}, 
	W^{(t)})(K_{\theta_y^{(t)}}(W^{(t)}) + g^{(t)}\mathbb{I}_n)^{-1} Y \label{eq:ymoments}
\\
\Sigma^{(t)}_Y(\mathcal{W}^{(t)}) &= \hat{\tau}^{2(t)}\left[K_{\theta_y^{(t)}}
	(\mathcal{W}^{(t)}) + g^{(t)} \mathbb{I}_{n'} -
	K_{\theta_y^{(t)}}(\mathcal{W}^{(t)}, W^{(t)}) (K_{\theta_y^{(t)}}(W^{(t)})
	+ g^{(t)}\mathbb{I}_n)^{-1}K_{\theta_y^{(t)}} (W^{(t)},
	\mathcal{W}^{(t)})\right]. \nonumber
\end{align}
Instead sampling $\tau^{2(t)}$ from its conjugate posterior conditional
$\mathrm{IG}(\frac{n}{2}, \frac{n\hat{\tau}^2}{2})$, and swapping in for
$\hat{\tau}^{2(t)}$ would be MVN.  In practice we find that this distinction
hardly matters, and it is sufficient to treat the former, in spite of estimated
scale $\hat{\tau}^{2(t)}$, as MVN except when $n$ is very small.  Although
posterior predictive uncertainty is most completely described by a large
collection of $\mathcal{Y}^{(t)}$, it can be far more compact to accumulate
moments through the law of total expectation and variance.  As long as the
samples $\mathcal{T}$ retained number more than $n'$,
\begin{equation}\label{eq:expmom}
\begin{aligned}
\mu_Y &= \frac{1}{|\mathcal{T}|}\sum_{t\in\mathcal{T}}\mu_Y^{(t)} & \mbox{and} &&
\Sigma_Y &= \frac{1}{|\mathcal{T}|}\sum_{t\in\mathcal{T}}\Sigma_Y^{(t)} + 
	\frac{1}{|\mathcal{T}|-n'} \sum_{t\in \mathcal{T}} (\mu_Y^{(t)} - \mu_Y) 
	(\mu_Y^{(t)} - \mu_Y)^\top,
\end{aligned}
\end{equation}
are moments depicting an MVN approximation to the empirical distribution 
$\{\mathcal{Y}^{(t)}\}_{t \in \mathcal{T}}$.

There are a few other shortcuts we find handy.  Unless sample paths of
$Y(\mathcal{X})$ are desired, one may short-cut a full covariance calculation
-- i.e., $\Sigma^{(t)}_Y(\mathcal{W}^{(t)})$ in Eq.~(\ref{eq:ymoments}) -- for
its diagonal (variance) components yielding point-wise equations
described by independent scalar Gaussian (or Student-$t$) equations. This
saves on both storage and computational effort, with the latter being
quadratic rather than cubic on $n'$.  Variances are sufficient for evaluating
AL criteria in Section \ref{sec:dgpal} and for calculating
quantile-based error-bars.  Such quantities would still capture all relevant
uncertainties for most purposes, e.g., for posterior predictive intervals.  
It may moreover be sufficient, and even desirable, to describe mean
uncertainty only, i.e., for $\mathbb{E}\{Y(\mathcal{X})\}$ rather than
$Y(\mathcal{X})$. Simply replace $g^{(t)}
\mathbb{I}_{n'}$ with the zero matrix in Eq.~(\ref{eq:ymoments}).  This is common
in AL via ALC and IMSE (targeting mean accuracy) and BO applications, and
when building ``confidence intervals'' on prediction. Note that the nugget is
still involved in $(K_{\theta_y^{(t)}}(W^{(t)}) + g^{(t)}\mathbb{I}_n)^{-1}$,
which is what induces smoothing.

Even after reducing to point-wise calculations in $Y(\mathcal{W}^{(t)})$,
extracting variances instead of covariances, the entire process is still cubic
in $n'$ because sampling $\mathcal{W}^{(t)}$ involves $n' \times n'$ matrices
(\ref{eq:wmoments}), which could be prohibitive.  However, for AL we find it
sufficient to skip calculating $\Sigma^{(t)}_{W_i}(\mathcal{X})$, and
subsequent MVN draws, and instead take $\mathcal{W}_i^{(t)} \equiv
\mu^{(t)}_{W_i}(\mathcal{X})$.  This results in an underestimate of predictive
uncertainty, but not substantially so and not in a way that has any detectable
effect on acquisitions.  In the running examples of Figures \ref{fig:ex} and \ref{fig:alc}, 
sampled $\mathcal{W}^{(t)}$ (as plotted in Figure \ref{fig:ess}) are indistinguishable 
from $\mu^{(t)}_{W}(\mathcal{X})$ and produce equivalent predictions/ALC/IMSE.

\subsection{Acquisition} \label{sec:acq}

Let $W_{n+1}^{(t)}$ denote the row-combined set of mapped latent layer values
for the existing data $W_n^{(t)}$ and those (transposed values) corresponding
to candidate/testing location $x_{n+1} \in \mathcal{X}$, commensurately $w^{(t)}_{n+1}
\in \mathcal{W}^{(t)}$, for a particular MCMC iteration $t\in \mathcal{T}$.  
We find it helpful to denote the un-scaled covariance as
$C_{n+1}^{(t)} = K_{\theta_y^{(t)}}(W_{n+1}^{(t)}) + g^{(t)}\mathbb{I}_n$,
where only the last row and column depend on $w^{(t)}_{n+1}$, and $C_n^{(t)}$ 
similarly.  Whereas $\mathcal{W}^{(t)}$ quantities play a fundamental role in the
input warping behind the predictive scheme from Section \ref{sec:pred},
$C_{n+1}^{(t)}$ evaluated point-wise for all $w_{n+1} \leftarrow x_{n+1} \in
\mathcal{X}$ is fundamental to ALC and IMSE acquisition under DGPs.

Consider IMSE first.  \citet{binois2019replication} explained how to integrate
predictive variance, i.e., $\Sigma_Y^{(t)}(w)$ in Eq.~(\ref{eq:ymoments}), in
closed form under a uniform measure in the unit cube. This is fine in ordinary
GP applications where $\mathcal{X}$ can be trivially coded to $[0,1]^d$, but
$w$'s are Gaussian and thus have support on the whole real line (in each of
$p$ coordinates).  Maintaining integration over a uniform measure, in keeping
with the spirit of earlier IMSE applications, thus required an upgrade to the
\citet{binois2019replication} development. Letting $a^{(t)} =
\min\{\mathcal{W}^{(t)}\}$, applied column-wise to calculate a $p$-vector, and
similarly $b^{(t)} = \max\{\mathcal{W}^{(t)}\}$, we derived the following
closed form IMSE over a uniform measure in  $[a^{(t)}, b^{(t)}]$ and Gaussian
kernel (\ref{eq:cov})
\begin{equation}
\label{eq:IMSE}
\mathrm{IMSE}(w_{n+1}^{(t)}) = \hat{\tau}^{2(t)}\prod_{i=1}^p(b_i^{(t)} - a_i^{(t)})
	\left[1 - \textrm{tr}\left(
	(C_{n+1}^{(t)})^{-1}H^{(t)}\right)\right]
\end{equation}
where $H^{(t)}$ is an $(n+1)\times(n+1)$ matrix with elements
\[
H^{(t)}_{jk} = \left(\frac{\pi\theta_y^{(t)}}{2}\right)^{\frac{p}{2}}\prod_{i=1}^p 
	\left\{\!\mathrm{exp}\left(
	-\frac{(w_{j,i} - w_{k,i})^2}{2\theta_y^{(t)}}\right) \left[\Phi\!\left(
	\frac{2b_i - w_{j,i} - w_{k,i}}{\sqrt{\theta_y^{(t)}}}\right) - \Phi\!\left(
	\frac{2a_i - w_{j,i} - w_{k,i}}{\sqrt{\theta_y^{(t)}}}\right)\right]\!\right\},
\]
in which $w_{j,i}$ is the $j^{th}$ element of the $i^{th}$ node of
$W_{n+1}^{(t)}$ and $\Phi$ is the standard Gaussian CDF.
A detailed derivation is provided in Supplement \ref{app:imse}.  IMSE is a
function of $w^{(t)}_{n+1}$, as the first $n$ rows of $W^{(t)}_{n+1}$ are
fixed.  The first $n\times n$ block of $H^{(t)}$ may be pre-calculated as only
the final row and column depend on $w^{(t)}_{n+1}$. Similarly, pre-calculating
$(C_n^{(t)})^{-1}$ using partitioned matrix inverse (Supplement
\ref{app:inv}), allows for linear computation of $(C_{n+1}^{(t)})^{-1}$ for
each $w^{(t)}_{n+1}$.

ALC proceeds similarly, but requires the specification of an additional
reference set $\mathcal{X}_{\mathrm{ref}}$.  We prefer
$\mathcal{X}_{\mathrm{ref}} \equiv \mathcal{X}$, which is the default in
other/ordinary GP contexts and simplifies matters cons{}iderably.  Although one
could map novel $\mathcal{X}_{\mathrm{ref}} \rightarrow
\mathcal{W}_{\mathrm{ref}}^{(t)}$ following Eq.~(\ref{eq:wmoments}), it is far
easier to take $\mathcal{W}_{\mathrm{ref}}^{(t)} = \mathcal{W}^{(t)}$.
We simplify notation by denoting $k_n^{(t)}(w) = K_{\theta_w^{(t)}}(W_n^{(t)}, w)$
and $k_{n+1}^{(t)}(w) = K_{\theta_w^{(t)}}(W_{n+1}^{(t)}, w)$.
Following Eq.~(\ref{eq:alc}), the ALC statistic may be developed as follows,
with details in Supplement \ref{app:alc}.
\begin{equation}
\begin{aligned}
\mathrm{ALC}(W_{n+1}^{(t)} \mid \mathcal{W}_{\mathrm{ref}}^{(t)}) &=
	\sum_{w\in \mathcal{W}^{(t)}_{\mathrm{ref}}} \hat{\tau}^{2(t)}k_{n+1}^{(t)\top}(w)
	(C_{n+1}^{(t)})^{-1}k_{n+1}^{(t)}(w) \\
	&\propto 
	\sum_{w\in \mathcal{W}_{\mathrm{ref}}^{(t)}} 
	\hat{\tau}^{2(t)} \left[v(h^\top k_n^{(t)}(w))^2 + 2zh^\top k_n^{(t)}(w) + 
	v^{-1}z^2 \right],
\end{aligned} \label{eq:alc_deriv}
\end{equation}
with $v = 1 + g^{(t)} - k_n^{(t)\top}(w_{n+1}^{(t)})(C_n^{(t)})^{-1}
	k_n^{(t)}(w_{n+1}^{(t)})$, 
$h = -v^{-1}(C_n^{(t)})^{-1}k_n^{(t)}(w_{n+1}^{(t)})$, and
$z = K_{\theta_y^{(t)}}(w_{n+1}, w)$.

Aggregate criteria (either IMSE or ALC) for $x_{n+1}$,  mapped
to $w_{n+1}^{(t)}$ for each $t\in\mathcal{T}$ arise as
\[
\mathrm{IMSE}(x_{n+1}) = \frac{1}{|\mathcal{T}|}\sum_{t\in\mathcal{T}} 
\mathrm{IMSE}(w^{(t)}_{n+1})
\quad\mathrm{and}\quad
\mathrm{ALC}(x_{n+1}) = \frac{1}{|\mathcal{T}|}\sum_{t\in\mathcal{T}} 
\mathrm{ALC}(w^{(t)}_{n+1}\mid \mathcal{W}^{(t)}_{\mathrm{ref}}).
\]
Depending on the preferred criterion, acquisition involves selecting either 
$x_{n+1} = \mathrm{argmin}_{x} \;\textrm{IMSE}(x)$ or 
$x_{n+1} = \mathrm{argmax}_{x} \;\textrm{ALC}(x)$.  These acquisitions could
be solved using a numerical optimizer, but we prefer search over a candidate set
because it lends well to pre-calculations, averaging over iterations, and parallel 
implementation.  More implementation details are provided in Section \ref{sec:implement}.

Extending our earlier illustrations, the right panel of Figure \ref{fig:ex}
shows IMSE and ALC evaluated over a dense candidate set $\mathcal{X}$.  Both
criteria favor acquisitions in the left, more wiggly region.  ALC's reference
set $\mathcal{X}_{\mathrm{ref}} =
\mathcal{X}$, which is uniform over the inputs, becomes warped through the
latent $W$ and yields non-uniform $\mathcal{W}_{\mathrm{ref}}$, accounting for
the slight differences between the IMSE and ALC surfaces.  This discrepancy
can become more pronounced in three-layer models, and higher.  An illustration
of the ALC surface for the 2d example of Section \ref{sec:sim} is provided
in supplementary material.
	
\section{Implementation and empirical evaluation} 
\label{sec:experiments}

After providing implementation details, we validate our DGP/AL methodology on
two hypothetical simulations and two real-world computer experiments.  To
benchmark against simpler alternatives out-of-sample, we evaluate root
mean-squared error (RMSE, smaller is better), and additionally a
proper scoring rule \citep[][Eq.~25]{gneiting2007strictly} proportional to
the predictive MVN log likelihood (larger is better) in the presence 
of noisy simulations.
We also compare the ability of the model to place design points
in the region of highest complexity, offering perhaps the most compelling
visual of sequential design success. We compare DGP models to both typical
stationary GP surrogates (using MCMC sampling of hyperparameters) and 
non-stationary treed GP models \citep[TGP;][]{gramacy2008bayesian} 
fit through the {\tt tgp} package on CRAN \citep{tgp}.
TGP uses trees to partition the input space, fits separate GPs on each
rectangular partition element, and averages all that across an MCMC posterior
sampling scheme.  We use the ALC selection criterion since it is 
already implemented in {\tt tgp} and allows for direct comparison.

\subsection{Implementation details} \label{sec:implement}

All empirical work in this manuscript is
supported by the {\tt deepgp} package on CRAN \citep{deepgp}.  Defaults,
described momentarily, are used throughout except where noted.  (For example,
we entertain $p < d$ in Section \ref{sec:drag}.)  These defaults are 
appropriate for inputs $X_n$
coded to $[0,1]^d$ and outputs $Y_n$ pre-processed to have mean zero and
variance one.  While the ESS component of Algorithm \ref{alg:gibbs} is free of
tuning parameters,  hyperparameter priors and proposals are required for
$\theta$ and $g$.  In the uniform sliding window proposal of
Eq.~(\ref{eq:proposal}), we use $l = 1$ and $u = 2$ for $g$ and $\theta$ at
all layers. 

Recall from Section \ref{sec:modelspec} that we take $\{\theta, g\}
\stackrel{\mathrm{iid}}{\sim} \mathrm{G}(3/2, b_{[\cdot]})$.  For $g$ we select
a rate so that 95\% of the prior mass lies below 1 (the scaled data variance),
resulting in $b_{[g]} = 3.9$.  Lengthscales $\theta$ proceed similarly, but we
differentiate between those which act on coded inputs $X_n$ spanning $[0,1]$,
those for latent layers under MVN spread, i.e., spanning $[-2,2]$ at 95\% (see
Figure \ref{fig:ess}), and latent depth.  In a three-layer model, $\theta_z$
acts on $X_n$, and we want to encourage an identity latent mapping in $Z$.
This helps regularize the system in the face of dynamics which are stationary,
nearly so, or for which we do not yet have enough data (say in an AL context)
to identify regime shifts. We thus choose $b_{[\theta_z]} = 3.9/4$, so that
95\% of the prior mass falls at less than four times the maximum pairwise
distances in $X_n$.  In a two-layer model, where $\theta_w$ acts on $X_n$, we
choose $b_{[\theta_w]}$  similarly. The outer layer always acts on latent
inputs in $[-2,2]$, but we are willing to substantially diverge from identity
mappings -- after all, presumably one entertains GPs because of an {\em a
priori} belief in nonlinearity. Following similar logic, but this time
targeting 95\% prior mass at less than 1.5 times the maximum pairwise distance
in $[-2,2]$, we choose $b_{[\theta_y]} = 3.9/6$. This choice is mirrored in
one-layer models, setting $b_{[\theta]} = 3.9/1.5$ for an isotropic
lengthscale acting on coded $X_n$.  Finally, in three-layer models we preserve
this hierarchy with $b_{[\theta_w]} = 3.9/12$ (95\% mass less than 3 times the
maximum pairwise distance).  These settings are, of course, all
user-adjustable.

Conducting MCMC sampling for two- and three-layer models
in the \texttt{deepgp} package using built-in defaults is as simple as the 
following, producing distinct S3-class objects. 
\begin{verbatim}
  R> fit.2 <- fit_two_layer(x, y)
  R> fit.3 <- fit_three_layer(x, y)
\end{verbatim}
Unless otherwise specified, the dimension of hidden layers will match that of
inputs, i.e., $p=d$.  For lower dimensional problems, like $d
\leq 3$ for our first three examples (Sections \ref{sec:sim}--\ref{sec:lgbb}),
it is important to allow for this flexibility, if only to imitate the kernel
structure of separable processes.  Smaller $p<d$ can have deleterious effects
except when the majority of (an already small number of) inputs are
irrelevant, which is rare in computer surrogate modeling applications. In
higher dimensional settings, upwards of $d=7$ as in Section \ref{sec:drag},
some tension between the dimension of inputs and layers can have a beneficial
``autoencoding effect'', uncovering a lower-dimensional latent
structure such as may arise when dynamics unfold on a sub-manifold of inputs. 

Once samples are collected, one may remove early iterations, saving burned-in
samples indexed as $t\in\mathcal{T}$.  To reduce the size of this set, samples
may further be thinned.  Both are facilitated by an S3 method called {\tt
trim}. When designing sequentially with AL, after new $\{x_{n+1},
f(x_{n+1})\}$ is acquired, it is helpful to re-start MCMC samples at the
values where the previous chain/model left-off, reducing the burn-in effort.
In our AL examples, we found that as few as 500 burn-in iterations were
sufficient.  Of course, the success of such shortcuts is intimately tied to
the rejection-free nature of ESS.  Efficient MCMC allows us to remove the
``human-in-the-loop'' in our AL exercises, specifying numbers of total and
burn-in iterations in advance, for all acquisitions, without the need to
investigate trace plots along the way.

After MCMC, which is an inherently serial affair, all downstream tasks are
parallelizable across iterations $t\in\mathcal{T}$.   Predictions may be made
separately for each $t$, with expectations (\ref{eq:expmom}) calculated
afterwards in a map--reduce fashion. When only point-wise means and
variances are required, predictions may be parallelized across testing
locations, $\mathcal{X}$.  AL criteria may similarly be parallelized across
iterations and across candidate locations -- one reason we embrace search over
a candidate set.  Parallel implementations of prediction, IMSE, and ALC are
provided by S3 methods {\tt predict}, {\tt IMSE}, and {\tt ALC} via
{\tt foreach} \citep{foreach} constructs.

\subsection{Simulated data} \label{sec:sim} 

\begin{figure}[ht!]
\centering
\includegraphics[width=18cm,trim=0 15 0 10]{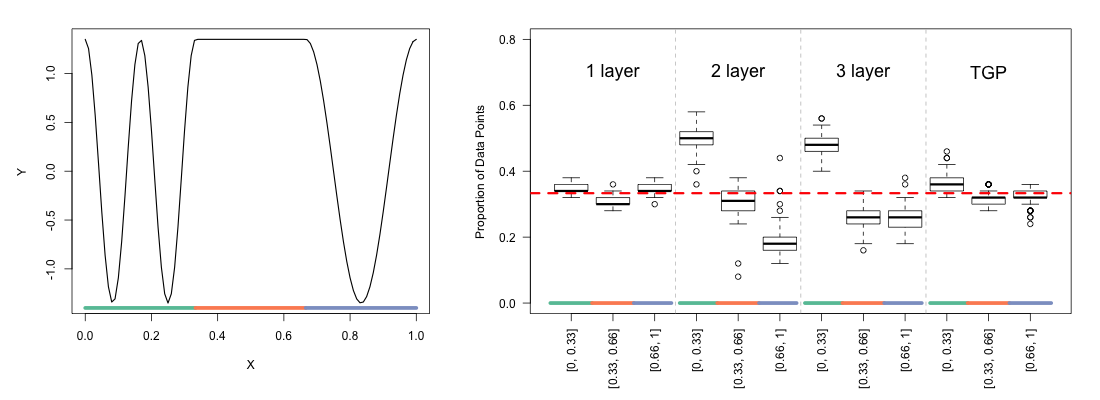}
\caption{{\em Left}: $f(x)$ from Eq.~(\ref{eq:oned}).  Three regions of the 
	input space are highlighted by the green, orange, and blue lines on the
	$x$-axis. {\em Right}:  proportion of sampled points falling in each region.
	Boxplots represent the spread of 100 repetitions.  A red dotted line marks
	the actual proportion of the input space occupied by each region.}
\label{fig:oned}
\end{figure}

\paragraph{1d.} Consider a 1d example generated piecewise:
\begin{equation}
\label{eq:oned}
f(x) = \begin{cases} 1.35 \cos(12\pi x) & x \in [0, 0.33] \\
					 1.35 & x \in [0.33, 0.66] \\
					 1.35 \cos(6\pi x) & x \in [0.66, 1]. \\
		\end{cases}
\end{equation}
Figure \ref{fig:oned} provides a visual.  The input space is divided into
three distinct, equally-sized regions.  We posit that $x\in[0, 0.33]$ is of
highest modeling interest because it is the most wiggly. Realizations are
observed with $\mathcal{N}(0,0.1^2)$ additive noise.  One-, two-, and
three-layer models were fit via AL acquiring a total $n=35$ runs initialized
via novel $n_0=10$ LHSs (i.e., 25 AL acquisitions) and 100-point
candidate/reference set(s) using ALC, all wrapped in a MC exercise with 100
repetitions.

\begin{figure}[ht!]
\centering
\includegraphics[width=18cm,trim=0 20 0 10]{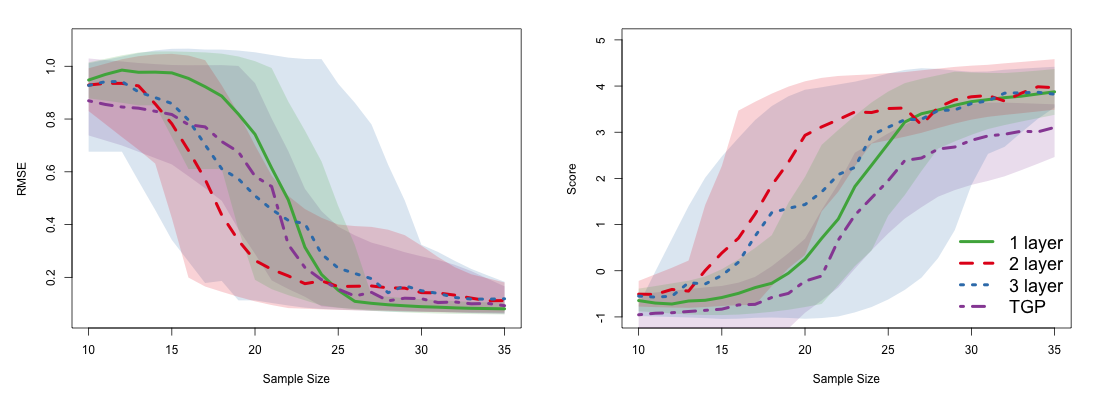}
\caption{RMSE (left) and score (right) for $f(x)$ from Eq.~(\ref{eq:oned}).  
	Solid lines represent the average over 100 repetitions.  Shaded regions 
	represent 90\% quantiles.}
\label{fig:onedresults}
\end{figure}

Figure \ref{fig:onedresults} shows RMSE and score calculated after each
iteration for the four comparators.  The two-layer DGP dramatically
outperforms the one-layer GP for small data sizes ($n=10,\dots,25$).  Once $n$
is large enough, the one-layer model catches up.  This ordinary GP eventually
edges out on RMSE, but the two-layer DGP wins on score. The latter's wider
RMSE quantiles at $n=35$ are caused by models that over-fit the noisy
observations.  The three-layer model performs well on average, but is marked
by very wide quantiles due to even more prominent over-fitting. Both DGPs do
a better job of placing runs in the region of highest interest, $x\in[0,
0.33]$; see Figure \ref{fig:oned}, right.  In the two-layer case, notice the
right-hand region $x
\in [0.66, 1]$ has the fewest acquisitions of all.  This happens, we believe,
because of the shape of the latent $W$ layer.  Refer again to the left panel
of Figure \ref{fig:ess}, where the right-hand region bends toward the
left-hand one, ``borrowing strength'' from the more substantial corpus of
information there.  Despite a strictly distance-based kernel structure,
such latent $W$ may impart a periodic effect on dynamics.  The non-stationarity 
of the TGP model allows some departure from space-filling designs, but 
performs poorly on score; likely a by-product of model uncertainty in the
spatial location of spits outlining partitions of the tree.  

\paragraph{2d.} Next consider a 2d example originally from
\citet{gramacy2009adaptive}:
\begin{equation}
\label{eq:twod}
f(x_1, x_2) = 10 x_1 \textrm{exp}\left(-x_1^2 - x_2^2\right) 
\quad\textrm{for}\quad x_1, x_2 \in [-2, 4],
\end{equation}
 observed with $\mathcal{N}(0, 0.1^2)$ noise. Although there are no abrupt
shifts, this function is characterized by two distinct regimes.  
\begin{figure}[ht!]
\centering
\includegraphics[width=18cm,trim=0 20 0 10]{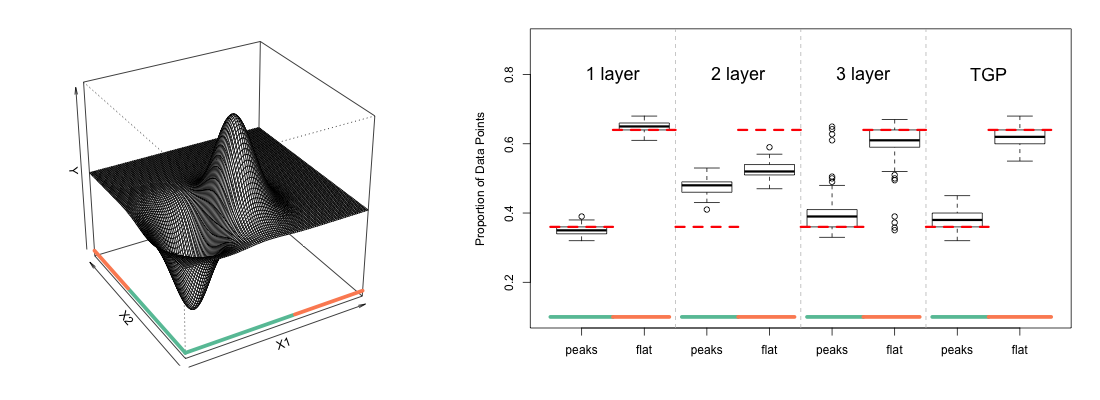}
\caption{{\em Left:} $f(x_1, x_2)$ from Eq.~(\ref{eq:twod}).  
  Two input regions are marked by the green and orange lines along the
	$x$-axes. {\em Right:} Proportion of AL acquisitions falling in each region.
	Boxplots indicate spread over 100 repetitions. Red dotted lines
	mark the actual proportions occupied by each region.}
\label{fig:twod}
\end{figure}
Figure
\ref{fig:twod} marks these with color along the $x$-axes; naturally
the ``peaky'', bottom-left region is of more interest than the much larger flat one.  
AL is initialized with random LHSs of size $n_0=10$, followed by 
ALC acquisitions (200 candidates/references) up to $n=80$.  
\begin{figure}[ht!]
\centering
\includegraphics[width=18cm,trim=0 20 0 10]{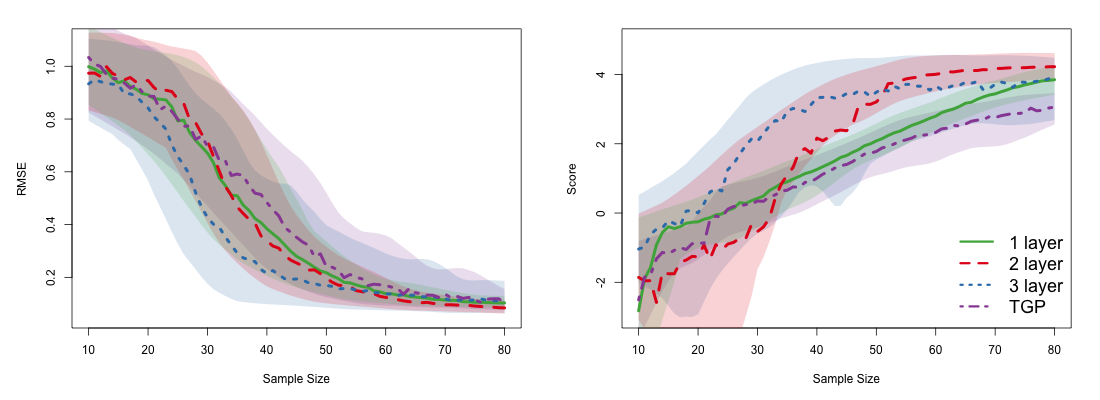}
\caption{RMSE (left) and score (right) for $f(x_1, x_2)$ from Eq.~(\ref{eq:twod}).  
	Solid lines represent the average over 100 repetitions.  Shaded regions represent 
	90\% quantiles.}
\label{fig:twodresults}
\end{figure}
Figure \ref{fig:twodresults} shows RMSE and score calculated after each
iteration for one-, two-, and three-layer models as well as TGP.  The two-layer 
DGP performs similarly to the one-layer GP in terms of RMSE, but the former achieves
consistently higher scores after a modest number of acquisitions ($n > 35$).
The three-layer DGP beats the others in both metrics on average but has the
widest quantiles. As in the 1d case, both deep models do a better job
of placing design points in the region of interest; see Figure \ref{fig:twod}.
TGP performs similarly to the one-layer GP.

\subsection{Langley Glide-Back Booster} 
\label{sec:lgbb}

The Langley Glide-Back Booster (LGBB) computer model was designed by NASA to
assess the movement of a rocket booster gliding back to Earth for re-use after
launching a payload into orbit.  A thorough review is provided in
\citet{pamadi2004aerodynamic}.  The LGBB simulator has three inputs
-- speed upon entry (mach), angle of attack (alpha), and side-slip angle (beta)
-- and produces six deterministic response values (lift, drag, pitch, side,
yaw, and roll).  This model is uniquely non-stationary because the sound
barrier at mach 1 imparts regime changes on aeronautic dynamics.

\citet{gramacy2008bayesian} developed Bayesian treed Gaussian process (TGP)
models specifically to account for these kinds of axis-aligned regime shifts.
In our work here, we utilize a
dense grid of TGP-surrogate evaluated LGBB response values from one of their AL
experiments, in lieu of real simulations (the actual simulator is
propriety).  Having the ``truth'' lie inside the TGP-model class elevates this 
model to gold standard in terms of out-of-sample validation, but nevertheless we find
comparison outside that class (with DGPs) to be informative.  For more details
on these data, and surrogate evaluations, see \citet[][Chapter
2]{gramacy2020surrogates}.

\begin{figure}[ht!]
\centering
\includegraphics[width=18cm,trim=0 10 0 15]{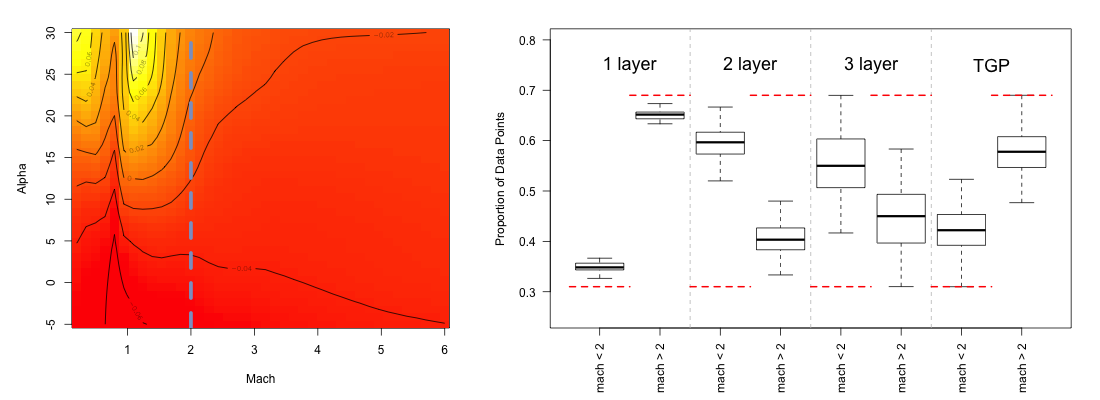}
\caption{{\em Left:} Visual of the LGBB side response for fixed beta = 4.  
	White/high, red/low.  A vertical dotted line splits the input region 
	at mach 2.  {\em Right:} Proportion of 300 
	sampled points falling left/right of mach 2.  
	Boxplots represent spread over 30 repetitions.  Red dotted lines mark 
	actual proportion of input space.}
\label{fig:rocket}
\end{figure}

Here we focus on the side response; dynamics in the other outputs are
similar. Figure \ref{fig:rocket} provides a visual of this response for a
fixed beta value. The sound barrier causes a sharp ridge at mach 1. The region
with mach values less than 2 is clearly the most ``interesting''.
After initializing with a random uniform (sub-) design size $n_0=50$ we
entertained ALC (500 candidates/references) for acquisitions up to $n=300$ 
using our usual four comparators.  A nugget of $g=10^{-8}$ was fixed to account for the 
deterministic nature of simulations.  RMSE on a 1000-point out-of-sample 
testing set (newly randomized for each calculation) was evaluated after every 
10th acquisition.  On those occasions, we re-used the 1000-point 
testing set as ALC candidates/references.  Joint evaluation of predictions and
ALC is efficient since both rely on 
mapped $\mathcal{W}^{(t)}$ (\ref{eq:wmoments}).

\begin{figure}[ht!]
\centering
\includegraphics[width=18cm,trim=0 20 0 10]{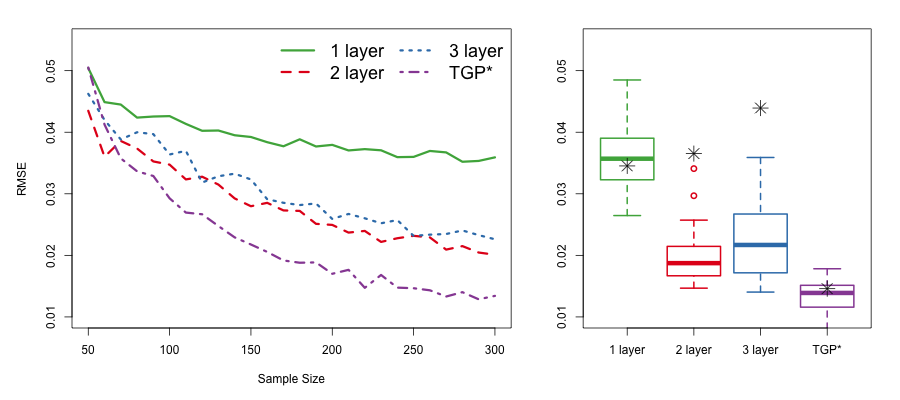}
\caption{{\em Left:} RMSE on out-of-sample 1000 point testing sets for the LGBB computer 
	experiment, averaged over 30 repetitions.  {\em Right:} spread of final RMSE
	at $n=300$ across 30 repetitions.  Black stars denote the RMSE obtained from
	static 300-point designs.  Both one-layer MLE variants performed
	equivalently to the one-layer MCMC fit.}
\label{fig:rresults}
\end{figure}

Figure \ref{fig:rresults} tracks out-of-sample RMSE
for the four comparators over AL acquisitions, via averages over thirty MC
repetitions (left), and the spread of values after the final AL acquisition
(right).  Both deep models outperform the one-layer GP model but are unable to
match the ``gold standard'' of TGP. The difference in the allocation of design
points among these models is pronounced.  See the right panel of Figure
\ref{fig:rocket}.  The one-layer model is very similar to a space-filling
design.  TGP deviates from space filling and puts more design points in the
region of interest (mach $< 2$), but still allocates the majority of points in
the flat region.  DGPs even more aggressively acquire/learn in the interesting
region, placing more than half of the inputs left of mach 2.

To further investigate the benefit of AL with DGPs, we generated a purely random
design of size $n= 300$ to train each of the four surrogates.  These RMSEs are
indicated as *s overlayed on the boxplots of their respective AL counterparts
in the right panel of Figure \ref{fig:rresults}.  Observe that these static
design results are substantially worse for the two- and three-layer fits.
These DGPs demand non-uniform acquisition to predict well. One-layer results
are comparable because these AL designs are essentially space-filling.  The
TGP results are curious.  Apparently, predictive prowess here comes from
``serendipitous accuracy'', arising from within-model simulations, more than
from AL reinforcement.  Finally, we fit a one-layer model using MLE
hyperparameters with both separable and non-separable lengthscales using the
\texttt{laGP} package \citep{laGP}.  These RMSEs are identical to those
obtained from the one-layer fit, with stars obscuring one another in the
figure.  We took this as in indication that the experiment was internally
well-calibrated.

\subsection{Satellite drag}
\label{sec:drag} 

Researchers at Los Alamos National Laboratory developed the \textit{Test
Particle Monte Carlo} simulator to compute drag coefficients for satellites in
low earth orbit. These are useful in the development of positioning and
collision avoidance systems.  \citet{sun2019emulating} created an \textsf{R}
wrapper for the simulator and made it publicly
available (\url{https://bitbucket.org/gramacylab/tpm/}). The simulator
relies on a geometric satellite specification, atmospheric composition, and 7
input variables. See \citet{sun2019emulating} and \citet[][Chapter
2]{gramacy2020surrogates} for a thorough discussion of the simulator and its
inputs.  We consider the GRACE satellite here; details on atmospheric
composition settings are available in our public Git
repository.  Each simulation takes about 86 seconds on a modern desktop.

\citet{mehta2014modeling} showed that over a restricted portion of the input
space, a GP surrogate trained via 1000-point LHS is able to predict drag
within 1\% root mean square percentage error (RMSPE).  We use Mehta's
work as a benchmark and show that sequentially designed DGP models can achieve
predictions within 1\% RMSPE with fewer data points.  We begin with a 7d
LHS of size $n_0=100$.  To help our model properly estimate the noise of the
stochastic simulator we ran (multiplicity two) replicates at a random
selection of ten of these original locations, so our training actually
comprised of $n_0 = 110$ runs.  To measure out-of-sample accuracy by RMSPE, 
and thus benchmark against \citeauthor{mehta2014modeling}, we built a single
1000-element testing set via LHS. 

\begin{figure}[ht!]
\centering
\includegraphics[width=18cm,trim=0 20 0 15]{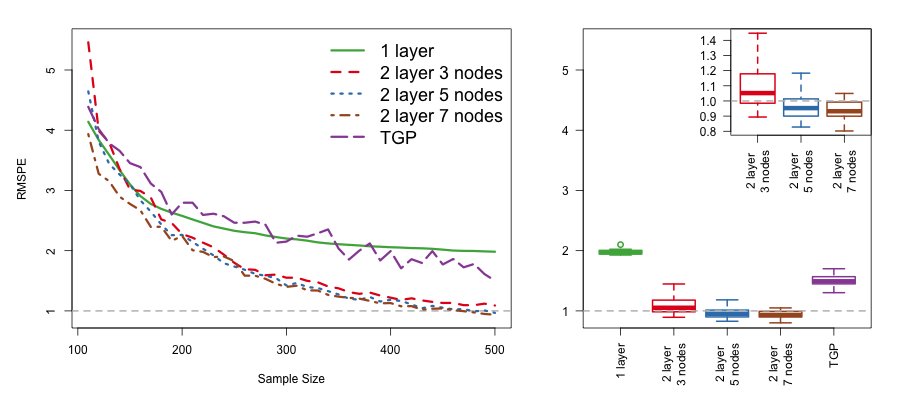}
\caption{{\em Left}: out-of-sample average RMSPE for sequential design of the
	satellite drag simulator over ten repetitions.  {\em Right}: spread of final
	RMSPE at $n=500$ across 10 repetitions.  Dashed lines highlight the 1\%
	goal.  The top-right sub-panel zooms in on the best comparators.}
\label{fig:satellite}
\end{figure}

Considering the substantial simulator expense, we did not entertain three-layer
DGPs.  Our previous experiments suggested that three-layer models were
overkill, and potentially risky.  Although their average-case behavior was
attractive, it was also more volatile over repetitions.  However, considering
the larger input dimension compared to those earlier exercises, we did
variations in latent dimension, with $p \in \{3, 5, 7\}$.  In all three cases,
training data inputs were sequentially selected using ALC 
(1000 candiates/references) up to $n=500$, and RMSPE on the testing set
was evaluated after every 10th acquisition. Again due to simulator expense, we
limited the MC exercise to ten repetitions (only five for TGP).
Results are summarized in Figure \ref{fig:satellite}.

Observe that all but the one-layer (ordinary GP) and TGP comparators were able
to achieve the 1\% benchmark, with high probability, with fewer than 500
runs of the simulator.   Interestingly, it is possible to accomplish this feat
with a latent dimension of $p=3 \ll d=7$.  Although results are improved with
more latent dimensions, up to $p=d=7$ which is the software default, this
comes at greater computational expense ($p=3$ is two times faster
on an eight core machine).

\section{Conclusion} 
\label{sec:conclude}

We provided a novel Gibbs-ESS-Metropolis method for fitting deep Gaussian
processes (DGPs), specifically targeted to surrogate modeling applications.
Their nonstationary nature makes them an excellent candidate for sequential
design via active learning (AL) when simulation dynamics are characterized by
stark regime shifts.  In addition to illustrations of component parts of our
inferential and AL scheme(s), we demonstrated how DGP models outperformed
typical GP regression in examples ranging up to seven dimensions.

Several modeling choices were made in order to protect identifiability and to
favor simplicity.  These were motivated by challenges brought to light in
empirical work behind earlier incantations.  Examples include enforcing common
dimensionality in latent layers, limiting depth to three-layers (or even two),
and limiting nodes to no more than the input dimension. These constraints seem
to provide a degree of regularization which does not overly hinder reactivity
of the apparatus, and appears to be essential in AL settings.  Early on, when
the training design is small, strong regularization helps encourage stability.
In later stages, after many runs have been acquired, the
information in the likelihood is able to dominate the prior. We
submit our favorable empirical outcomes, on a relatively broad range of
synthetic and real-simulator AL examples, as evidence.  However, a more
detailed cost-benefit analysis of DGP complexity v.~prowess may be
warranted.

We were most surprised by how well a simple default setup -- two layers with
nodes matching the input dimension $(p=d)$ -- compared to more complex
alternatives. Although there is evidence that even deeper GPs ($\geq 3$ latent
layers) work well when there are abrupt regime shifts
\citep{damianou2013deep,dunlop2018how}, we were unable to identify any common
computer simulation benchmarks which demanded that complexity.  It may be that
typical simulators simply aren't that pathological. Perhaps incorporating
input connected networks \citep{duvenaud2014avoiding}, in which deep layers
depend on both the previous layer and the original inputs, may increase the utility
of deeper models.  In higher dimensional settings, we found the relative
success of two-layer DGPs with $p < d$ to be intriguing.   Our work was limited to
post-hoc analysis of depth and node size; selecting these specifications in
real-time, for example, would be of great practical interest.

The recent success of DGPs as surrogates, such as for calibration
\citep{marmin2018variational}, multi-fidelity modeling
\citep{cutajar2019deep}, and black-box optimization
\citep{hebbal2021bayesian}, is exciting.  UQ plays a key role in each of
these, yet approximation via maximization is the {\em modus operandi}.
Consequently, such surrogates under-estimate -- or at best mis-characterize --
relevant uncertainties.  Our novel Gibbs-ESS-Metropolis scheme may make MCMC
tractable where it wasn't previously effective, and the additional UQ may
result in significant improvements.

The additional computation required to fit DGP models poses a challenge for
larger sample sizes.  Our MCMC method is not appropriate when data is
abundant.  Potential remedies include incorporating inducing points
\citep{rajaram2021empirical} and/or utilizing the covariance function or
operator formulations of \citet{dunlop2018how}.  Moreover, it may
be possible to implement sequential model updates in our framework following
\citet{wang2016sequential}.

\subsection*{Acknowledgements}

This work was supported by the U.S. Department of Energy, Office of
Science, Office of Advanced Scientific Computing Research and Office of
High Energy Physics, Scientific Discovery through Advanced Computing
(SciDAC) program under Award Number 0000231018.
We also thank referees for their valuable insights and suggestions 
that led to an improved manuscript.

\bigskip
\begin{center}
{\large\bf SUPPLEMENTARY MATERIAL}
\end{center}

\begin{description}

\item[] Illustration of ALC surface for the 2d example
	and derivations of IMSE (\ref{eq:IMSE}) and ALC (\ref{eq:alc_deriv}).
\end{description}

\bibliographystyle{jasa}
\bibliography{references}

\newpage
\begin{center}
{\large\bf SUPPLEMENTARY MATERIAL}
\end{center}
\appendix

\section{ALC Illustration}

\subsection{2d ALC Surface}

\begin{figure}[ht!]
\centering
\includegraphics[width=12cm]{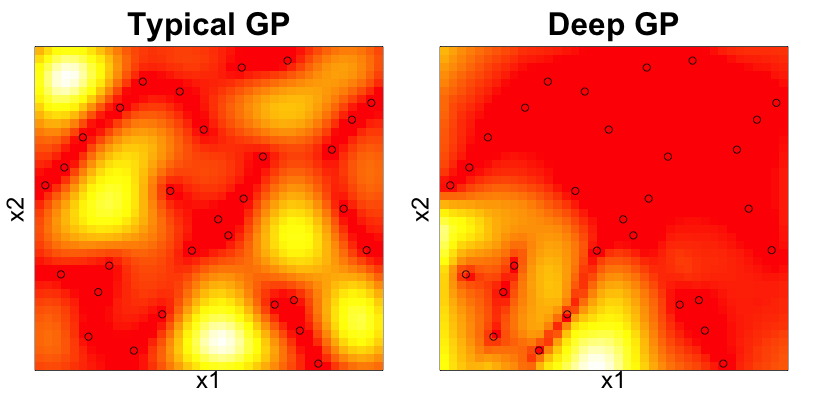}
\caption{Heat map of ALC surface for typical GP and two-layer DGP
	surrogates (via MCMC) for $f(x_1, x_2)$ in Eq.~(\ref{eq:twod}).  Open circles 
	indicate design $X_n$.  White/high, red/low.}
\label{fig:alc}
\end{figure}

For another illustration, consider a typical, stationary GP surrogate fit 
to simulations following Eq.~(\ref{eq:twod}), detailed in Section 
\ref{sec:sim}. The left panel of Figure \ref{fig:alc} shows ALC, using a 
dense $X_{\mathrm{ref}}$ grid, after training on evaluations obtained 
from an LHS of size $n = 30$.  In this example, the lower left corner 
is the area of highest interest, but ALC is, in essence, a function of 
proximity to $X_n$.  A two-layer DGP fit to the same training data
is able to allocate uncertainty in the ``interesting'' region, resulting 
in non-space filling acquisitions.

\section{Derivations}

\subsection{Partitioned Matrix Inverse}
\label{app:inv}

Following the properties of partitioned matrices
\citep[e.g.,][]{barnett1979matrix}, with $k_n(w_{n+1}) = K_{\theta_y}(W_n,
w_{n+1})$,
\[C_{n+1} = \begin{bmatrix} C_n & k_n(w_{n+1})\\
	k_n^\top(w_{n+1}) & 1 + g
\end{bmatrix} 
\quad\mathrm{yields}\quad
C_{n+1}^{-1} = \begin{bmatrix} \left[C_n^{-1} + h h^\top v\right] & h \\
	h^\top & v^{-1}
\end{bmatrix}
\]
where $v = 1 + g - k_n^\top(w_{n+1})C_n^{-1} k_n(w_{n+1})$ and 
$h = -v^{-1}C_n^{-1}k_n(w_{n+1})$.

\subsection{IMSE Derivation}
\label{app:imse}

Here we extend the closed-form IMSE derivation of
\citet{binois2019replication} to allow integration under uniform measure over
the general domain $W_i \in [a_i, b_i]$ for $i = 1, \dots, p$ for the model
$Y\sim N_n(0, K_{\theta_y}(W) + g\mathbb{I})$ and isotropic Gaussian kernel
(\ref{eq:cov}).  Application to MCMC samples requires indexing with iteration
$(t)$, but we drop this notation here for convenience. We set $\Sigma(w) = 1$
instead of $\Sigma(w) = 1 + g$ to target the variance of the mean.

Using $W_{n+1}$ and $C_{n+1}$ as described in Section \ref{sec:acq}, and additionally 
denoting $k_{n+1}(w) = K_{\theta_y}(W_{n+1}, w)$, 
IMSE may be expressed as follows.
\begin{align*}
\textrm{IMSE}(W_{n+1}) &= \int_{a_p}^{b_p}\dots\int_{a_1}^{b_1} \Sigma_Y(w) \;dw_1\dots dw_p \\
	&= \prod_{i=1}^p(b_i - a_i) \mathbb{E}\left[\Sigma_Y(w)\right] \\
	&= \prod_{i=1}^p(b_i - a_i) \mathbb{E}\left[\hat{\tau}^2\left(1 - k^\top_{n+1}(w) 
		C_{n+1}^{-1} k_{n+1}(w)\right)\right] \\
	&= \hat{\tau}^2 \prod_{i=1}^p(b_i - a_i)\left[1 - \mathbb{E}\left[k^\top_{n+1}(w) 
		C_{n+1}^{-1} k_{n+1}(w)\right]\right]
\end{align*}
This expectation reduces to a trace of matrix products following Lemma 3.1 
of \citet{binois2019replication},
\[
\mathbb{E}\left[ k^\top_{n+1}(w) C_{n+1}^{-1} k_{n+1}(w) \right] = 
	\mathrm{tr}\left(C_{n+1}^{-1} H \right),
\]
where the elements of $H$ are defined as
\begin{align*}
H_{jk} &= \prod_{i=1}^p \int_{a_i}^{b_i} K_{\theta_y}(w_{j,i}, w)
			K_{\theta_y}(w_{k,i}, w) \;dw_i \\
	&= \prod_{i=1}^p \int_{a_i}^{b_i} \mathrm{exp}\left(\frac{-(w_{j,i} - w)^2}{\theta_y}\right)
		\mathrm{exp}\left(\frac{-(w_{k,i} - w)^2}{\theta_y}\right) \;dw_i\\
	&= \prod_{i=1}^p \int_{a_i}^{b_i} \mathrm{exp}\left(-\frac{2}{\theta_y}
		\left(w - \frac{w_{j,i} + w_{k,i}}{2}\right)^2
		- \frac{1}{2\theta_y}(w_{j,i} - w_{k,i})^2\right)\;dw_i \\
	&= \prod_{i=1}^p \left[\mathrm{exp}\left(-\frac{(w_{j,i} - w_{k,i})^2}{2\theta_y}\right) 
		\int_{a_i}^{b_i}\mathrm{exp}\left(-\frac{2}{\theta_y}
		\left(w - \frac{w_{j,i} + w_{k,i}}{2}\right)^2\right)\;dw_i \right]\\
	&= \prod_{i=1}^p \left[\sqrt{\frac{\pi\theta_y}{2}}\mathrm{exp}
		\left(-\frac{(w_{j,i} - w_{k,i})^2}{2\theta_y}\right) \int_{a_i}^{b_i}
		\mathcal{N}\left(w \mid \mu = \frac{w_{j,i} + w_{k,i}}{2}, \sigma^2 = \frac{\theta_y}{4}\right) 
		\;dw_i \right] \\
	&= \left(\frac{\pi\theta_y}{2}\right)^{\frac{p}{2}} \prod_{i=1}^p 
		\left\{\mathrm{exp}\left(-\frac{(w_{j,i} - w_{k,i})^2}{2\theta_y}\right) 
		\left[\Phi\left(\frac{2b_i - w_{j,i} - w_{k,i}}{\sqrt{\theta_y}}\right) 
		- \Phi\left(\frac{2a_i - w_{j,i} - w_{k,i}}{\sqrt{\theta_y}}\right)\right]\right\}.
\end{align*}
Above, $w_{j,i}$ is the $j^{th}$ element of the $i^{th}$ node of 
$W$ and $\Phi$ is the standard Gaussian cumulative distribution function (CDF).

\subsection{ALC Derivation}
\label{app:alc}

Here, we detail the re-expression of ALC from Eq.~(\ref{eq:alc_deriv}) in terms of
the new latent element, $w_{n+1}$.  Again, we drop $(t)$ indexing to
streamline. Quantities $k_n(w)$, $v$, and $h$ are defined in Appendix
\ref{app:inv} and yield partitioned representation of $k_{n+1}^\top(w) =
\begin{bmatrix} k_n^\top(w) \mid z \end{bmatrix}$ where $z =
K_{\theta_y}(w_{n+1}, w)$.  Tedious matrix multiplication gives
\begin{align*}
k_{n+1}^\top(w)C_{n+1}^{-1}k_{n+1}(w)
	&= 
	\begin{bmatrix} k_n^\top(w) \mid z \end{bmatrix}
	\begin{bmatrix} C_n^{-1} + hh^\top v & h \\  h^\top & v^{-1} \end{bmatrix}
	\begin{bmatrix} k_n(w) \\ z \end{bmatrix} \\
	&= \begin{bmatrix} k_n^\top(w)\left(C_n^{-1} + hh^\top v\right) + zh^\top \mid
			k_n^\top(w)h + zv^{-1} \end{bmatrix} 
		\begin{bmatrix} k_n(w) \\ z \end{bmatrix} \\
	&= k_n^\top(w)(C_n^{-1} + hh^\top v)k_n(w) + zh^\top k_n(w) + 
		k_n^\top(w)hz + v^{-1}z^2 \\
	&= k_n^\top(w)C_n^{-1}k_n(w) + v(k_n^\top(w)h)^2 + 2zk_n^\top(w)h + v^{-1}z^2 \\
	&\propto v(k_n^\top(w)h)^2 + 2zk_n^\top(w)h + v^{-1}z^2,
\end{align*}
producing the following ALC 
\[
\mathrm{ALC}(w_{n+1}\mid \mathcal{W}_{\mathrm{ref}}) \propto
\sum_{w\in\mathcal{W}_{\mathrm{ref}}}
\hat{\tau}^2\left[v(h^\top k_n(w))^2 + 2zh^\top k_n(w) + v^{-1}z^2\right].
\]

\end{document}